# Pressure dependent DFT study of mechanical, thermal, electronic, magnetic, and optical properties of $EuCd_2As_2$: a magnetic semimetal

Salauddin Muhammad Anis, M. A. Afzal, R. S. Islam*, S. H. Naqib*

Department of Physics, University of Rajshahi, Rajshahi 6205, Bangladesh

Corresponding Authors, Emails: rsislam@ru.ac.bd (R. S. Islam) and jsr_eda1@ru.ac.bd (S. H. Naqib)

## Abstract

We present a comprehensive spin-polarized density functional theory (DFT) investigation of the pressure-dependent physical properties of $EuCd_2As_2$, a magnetically and topologically rich compound with potential applications in spintronics and optoelectronics technologies. We have investigated the structural, mechanical, thermophysical, acoustic, electronic, magnetic, and optical properties of this material at pressures of 0, 2, 4, 6, and 8 GPa. The calculated lattice parameters at zero pressure are in good agreement with experimental values, which validates the reliability of the present computational approach. The mechanical stability of $EuCd_2As_2$ is maintained over the investigated pressure range. The elastic constants, bulk modulus, shear modulus, and Young's modulus are found to be tunable with applied pressure. The compound undergoes a brittle to ductile transition at higher pressures. The Debye temperature, melting temperature, and Grüneisen parameter are found to be strongly pressure-dependent. Phonon thermal conductivity has been evaluated. The electronic properties of $EuCd_2As_2$ were studied through band structure and density of states (DOS). The effect of pressure up to 8 GPa was investigated for the ferromagnetic (FM) phase. At 0 GPa, the calculated FM band structure exhibits no band gap at the Fermi level, which confirms the semimetallic nature of $EuCd_2As_2$. The net magnetic moment of 6.96 $\mu_B$ per unit cell was obtained for this compound indicating its potential for spintronics application. The optical properties, including the dielectric function, refractive index, absorption coefficient, optical conductivity, reflectivity, and loss function, are calculated. The calculated optical spectra indicate strong absorption in the visible and ultraviolet regions, suggesting that $EuCd_2As_2$ may have potential for optoelectronic applications.

## 1. Introduction

$EuCd_2As_2$ has emerged as a compelling system to study the interplay between magnetism and band topology [1]. This compound has a trigonal crystal structure with a space group $P\bar{3}m1$, in which triangular Eu layers are separated by $Cd_2As_2$ bilayers, with magnetism arising primarily from the localized $Eu^{2+}$ ($4f^7$, S = 7/2) [2]. The low-temperature ground state of $EuCd_2As_2$ has an interesting feature, namely, its electronic properties can change depending on the behavior of its magnetic order at low temperatures. Jo et al. [1] have reported that, depending on the level of band filling, controlled by the Eu-site occupancy during crystal growth, the material can be either in an antiferromagnetic phase or in a ferromagnetic phase. Comprehensive density functional theory (DFT) calculations indicate that the antiferromagnetic (AFM) state can host topological states. In contrast, the ferromagnetic (FM) state exhibits two pairs of Weyl points tuned to one pair of Weyl points by polarizing the spins along the crystallographic c-direction [3]. Roychowdhury et al. [4] synthesized the FM phase of $EuCd_2As_2$ using the salt flux method. They reported the contribution of the Nernst effect and anomalous Hall conductivity in this material caused by the Berry curvature. Moreover, they found that magnetic-field-induced spin canting modifies the electronic band structure and the associated topological response [4]. Ma et al. [5] investigated the temperature-dependent electronic band structure study. They have found the Weyl nodes in the paramagnetic phase below 100 K that originate due to the interaction between the itinerant electrons and localized moments in the paramagnetic phase. However, above approximately 100 K, they reported a crossover from a Weyl-semimetal-like state toward a Dirac-semimetal-like electronic structure. Rahn et al. [6] have synthesized the A-type antiferromagnetic phase of $EuCd_2As_2$, with Eu spins aligned along the ab plane. In this phase, (DFT) calculations suggest that the $C_3$ symmetry is broken, resulting in the opening of a band gap in the vicinity of the Fermi surface. Conversely, when the spins are oriented along the (c)-axis in the antiferromagnetic configuration, the compound is predicted to exhibit a Dirac-semimetal-like state with no gap at the crossing point [6]. However, recent magneto-infrared spectroscopy studies have given new insights into the magneto-optical properties of $EuCd_2As_2$. D. Santos-Cottin et al. [7] have identified $EuCd_2As_2$ as a magnetic semiconductor, challenging the previous study of identifying $EuCd_2As_2$ as a topological semimetal. While ab initio calculations had suggested Weyl points, and angle-resolved photoemission spectroscopy (ARPES) measurements had revealed a conical shape of the valence band, infrared spectroscopy revealed that $EuCd_2As_2$ hosts a significant band gap of 0.77 eV. This finding is particularly significant as several recent studies reproduced $EuCd_2As_2$ samples with activated resistivity behavior and confirmed the presence of a substantial band gap. In a magnetic field, the energy gap is modified, but the band gap still remains even in high magnetic fields (16 T), demonstrating a strong magneto-optical response [8]. All these findings suggest that $EuCd_2As_2$ as a particularly interesting platform for investigating the interplay between magnetism, electronic structure, and topology, making it a promising material for developing a fundamental understanding of physics as well as for technological applications such as optoelectronics and spintronics. So far, several theoretical investigations have been carried out on $EuCd_2As_2$, such as the electronic configuration, magnetic characteristics, electrical conductivity, and optical attributes of $EuCd_2As_2$ [9]. Thermoelectric parameters such as Seebeck coefficient, electrical and thermal conductivity, power factor, and figure of merit (ZT) as a function of temperature, as well as chemical potential of $EuCd_2As_2$ for anti-ferromagnetic phase, have been investigated using ab-initio (DFT) and semi-classical Boltzmann transport theory within constant relaxation time approximation [10]. Moreover, DFT calculations have been performed to analyze electronic band structure of $EuCd_2As_2$ for different magnetic configurations, including in-plane, out-of-plane, and spin-canted states [11]. Additionally, it was analyzed how spin orientation and canting affect the formation and position of Weyl points near the Fermi level [11].

To our knowledge, a systematic first-principles investigation of its pressure-dependent elastic, mechanical, thermophysical, acoustic, and optical properties remains limited for $EuCd_2As_2$. Pressure is a clean tunning parameter that can be varied continuously to control the physical properties of solids. In this work we have investigated a large number of electronic and thermophysical properties of $EuCd_2As_2$ within a moderate pressure range from 0 to 8 GPa. This range is easily accessible in high pressure laboratories. Our findings not only show a novel pressure dependent pathway to optimize the material's mechanical stability and thermal management but also hold promise for increasing its efficiency in spintronic and thermoelectric applications. Taddei et al. [11], [12] experimentally and computationally studied a FM $EuCd_2As_2$. In the present work, we extend the study by Taddei et al. by investigating pressure-dependent structural, elastic, mechanical, acoustic, thermophysical, electronic, and optical properties of the FM phase $EuCd_2As_2$ using first-principles calculations. This work presents a comprehensive property coverage by combining structural, mechanical, electronic, optical, and thermal properties, providing a broad-based understanding of the material.

## 2. Computational methodology

The structural, elastic, thermo-physical, magnetic, and optical properties of $EuCd_2As_2$ were studied using density functional theory (DFT) within the generalized gradient approximation (GGA) [13]. The calculations were carried out using the Perdew–Burke–Ernzerhof (PBE) exchange-correlation functional as implemented in the CASTEP code [14]. The ultrasoft pseudopotential setting is used for this compound. For the CASTEP calculations, a plane-wave cut-off energy of 750 eV and a Monkhorst–Pack k-point grid [15] of 12×12×7 were used to sample the Brillouin zone (BZ). Since $EuCd_2As_2$ is a magnetic material, spin-polarized calculations with collinear treatment were performed. The crystal structure was optimized using the BFGS minimization method [16]. During this calculation the total energy convergence tolerance was set to $5.0\times10^{-6}$ eV/atom, while the maximum ionic force, maximum ionic displacement, and maximum stress were limited to 0.01 eV/Å, $5.0\times10^{-4}$ Å, and 0.02 GPa, respectively. The independent single-crystal elastic constants ($C_{ij}$), bulk modulus (B), and shear modulus (G) were calculated using the stress–strain method implemented in the CASTEP code. After obtaining the optimized structure, the electronic and magnetic properties were calculated. The magnetic behavior was analyzed from self-consistent spin-polarized calculations. The optical properties were obtained from the frequency-dependent dielectric function. Other optical parameters were evaluated through the Kramers–Kronig relations available in the CASTEP package. The optical properties of materials can be understood by their frequency dependent dielectric function $\epsilon(\omega)$. The dielectric function can be written as $\epsilon(\omega) = \epsilon_1(\omega) + i\epsilon_2(\omega)$. Here $\epsilon_1(\omega)$ and $\epsilon_2(\omega)$ correspond to the real and imaginary parts of the dielectric constants, respectively. For electronic band structure Quantum ESPRESSO [17] GGA approximation with PBE functional is used. Since $EuCd_2As_2$ is a magnetic material, spin-polarized calculations were carried out during the study. Interatomic interactions and electronic structures were described using a scalar relativistic pseudopotentials sourced from the Standard Solid State Pseudopotentials (SSSP) library within the PBE generalized gradient approximation and [18]. A plane-wave kinetic energy cutoff of 55 Ry and a charge density cutoff of 440 Ry were used in all calculations. A Monkhorst–Pack k-point grid [15] of 12×12×7 were used to sample the Brillouin zone (BZ). The self-consistent field calculations were performed with an energy convergence threshold of $4 \times 10^{-5}$ Ry. And the force convergence threshold was $5 \times 10^{-4}$ Ry/Bohr. Charge density mixing was carried out using the plain mixing scheme with a mixing parameter of 0.3.

## 3. Results and discussion

### 3.1 Structural Properties

The compound $EuCd_2As_2$ has a trigonal crystal structure with a specific space group. Almost all physical properties, such as band structure, optical properties, and elastic constants, depend on the crystal structure and the nature of the constituent atoms. The crystal structures of $EuCd_2As_2$ are shown in **Figs. 1 and 2**. These crystal structures were visualized using VESTA [19]. The Wyckoff positions are taken from a previous study [2]. The $EuCd_2As_2$ has a layered structure. In this structure the $Cd_2As_2$ bilayers are separated by the triangular Eu layers. The magnetic moments primarily originates from the $Eu^{2+}$ ions [5]. The lattice parameters obtained from optimization at 0 GPa were compared with those of experimental results in **Table 1**. It is seen from **Table 1** that the calculated lattice parameters a = b and c for $EuCd_2As_2$ are in good agreement with the experimental results. Moreover, the volume of the unit cell is very close to the experimental result. This is a crucial step to judge the reliability of this computational work. Moreover, lattice parameters are studied under several pressures up to 8 GPa, and the result is shown in **Fig. 3** and **Table 1**. The lattice parameters are found to decrease with increasing pressure. The unit cell volume decreased by 11.70% as pressure increased from 0 GPa to 8 GPa. In the **Fig. 3** $a_0$, $c_o$ $and$ $V_o$ are the lattice parameters and unit cell volume respectively under zero pressure.

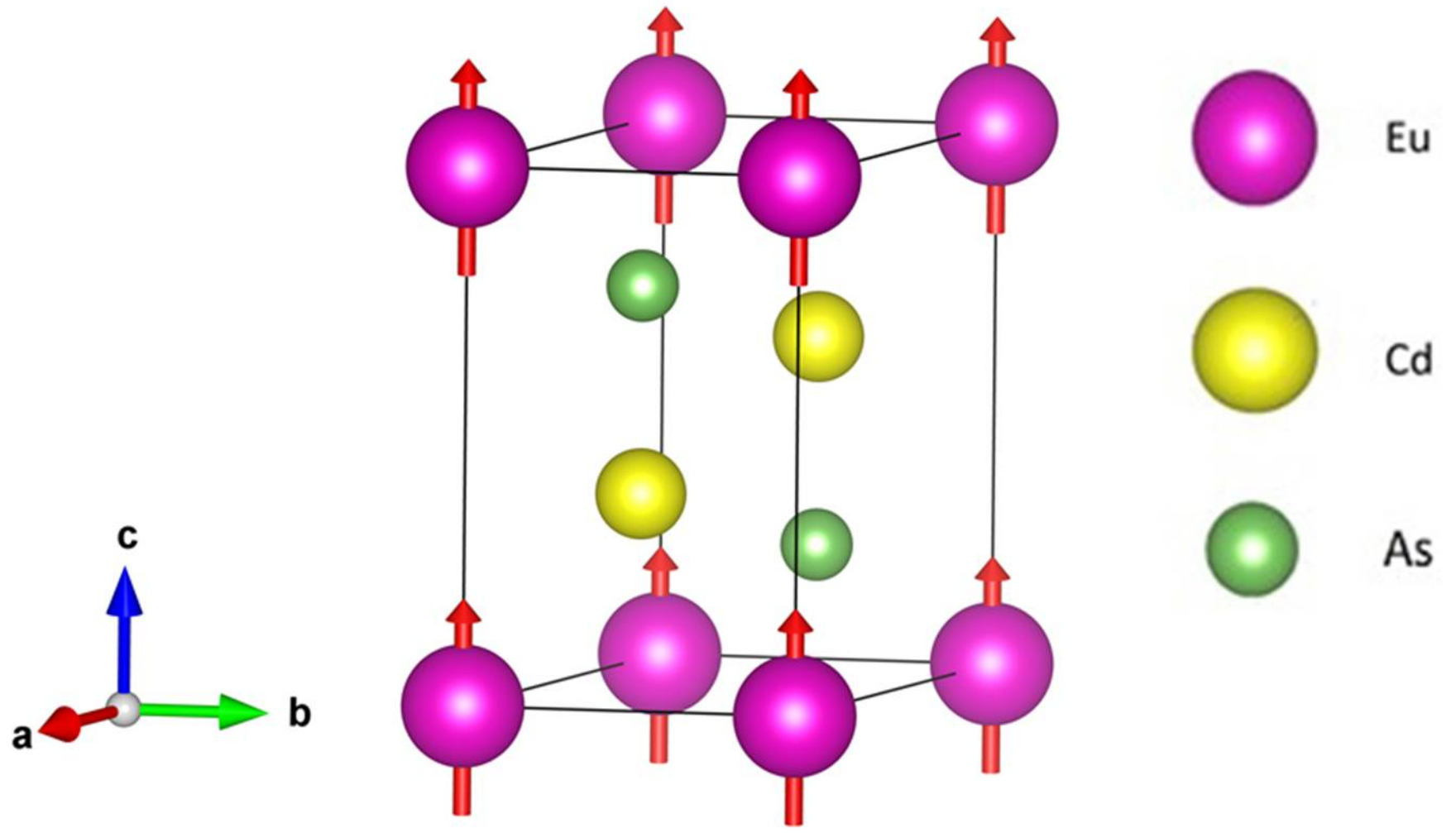

**Figure 1.** Three-dimensional (3D) crystal structure of $EuCd_2As_2$ where the vectors indicates the direction of magnetic moment.

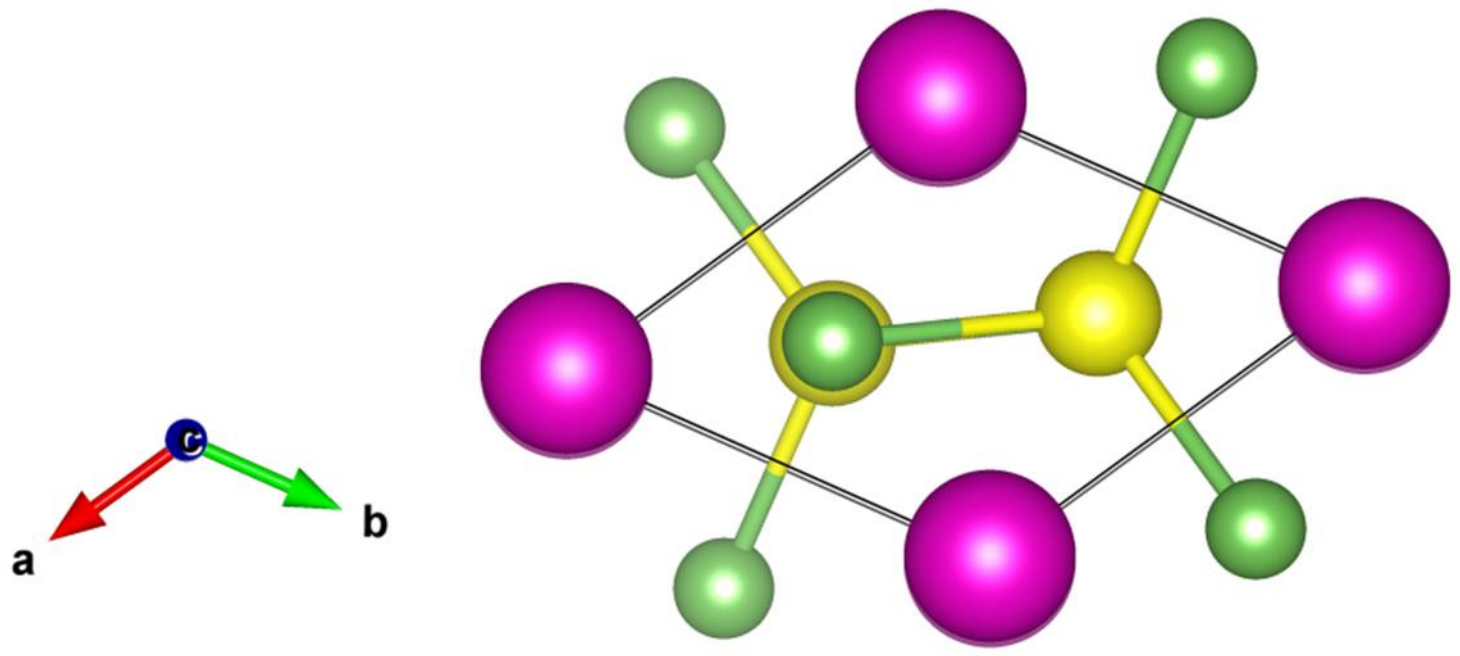

**Figure 2**. Two-dimensional 2D view of the ab-plane of crystal structure of $EuCd_2As_2$.

**Table 1**. Optimized lattice parameters under pressures (a = b, and c in Å), cell volume V in Å$^3$ of $EuCd_2As_2$.

| Pressure (GPa) | a = b | c | $a/a_0$ | $c/c_0$ | V | $V/V_o$ | Remarks |
|---|---|---|---|---|---|---|---|
| - | 4.450 | 7.350 | - | - | 126.000 | - | [2] |
| 0 | 4.500 | 7.322 | 1 | 1 | 128.405 | 1 | This work |
| 2 | 4.452 | 7.203 | 0.989 | 0.983 | 123.623 | 0.963 | |
| 4 | 4.412 | 7.102 | 0.980 | 0.970 | 119.693 | 0.932 | |
| 6 | 4.376 | 7.013 | 0.972 | 0.958 | 116.326 | 0.906 | |
| 8 | 4.345 | 6.934 | 0.966 | 0.947 | 113.388 | 0.883 | |

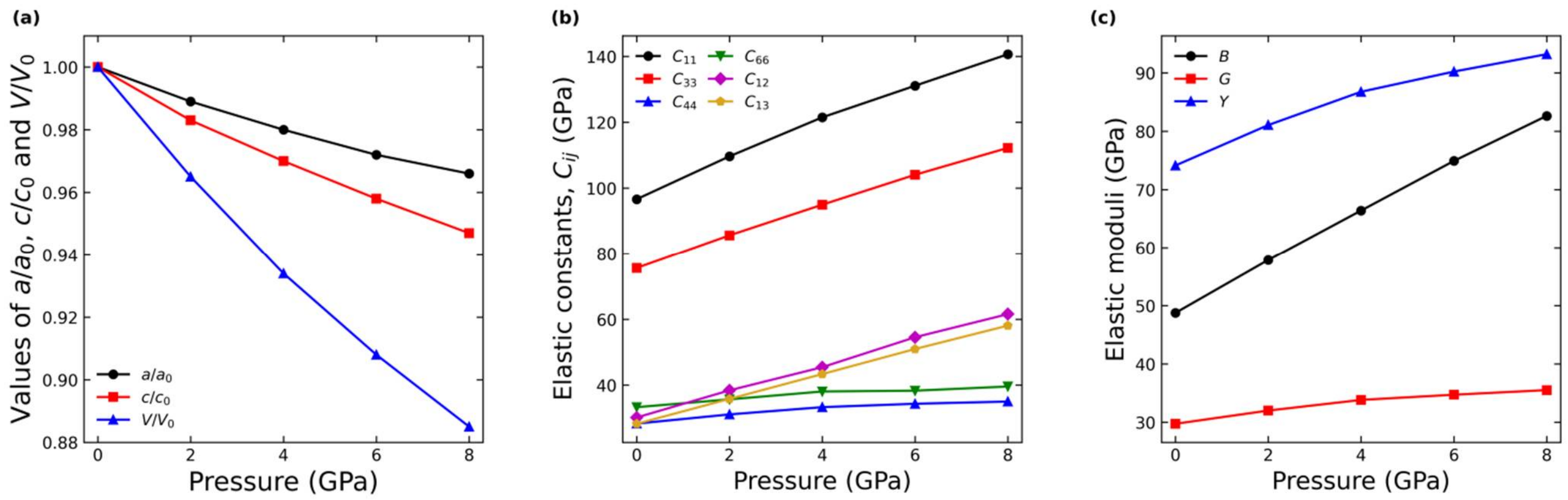

**Figure 3.** (a) Normalized lattice parameters and volume ($a/a_o$, $c/c_o$, $V/V_o$), b) Elastic constants $C_{ij}$ and, (c) Elastic moduli under pressures.

## 3.2. Elastic Properties

### 3.2.1 Elastic constants

From the elastic constants of a crystalline solid, the mechanical behavior and stability under external stress can be understood. These elastic constants are related to the material's mechanical stability, stiffness, ductility, brittleness, and elastic anisotropy [20]. For a material to be mechanically stable, the Born-Huang conditions must be satisfied. As the selected compound $EuCd_2As_2$ has a trigonal crystal structure, stability conditions under pressure are [21]:

$$C_{44} - P > 0$$

$$C_{11} - P - |C_{12} + P| > 0$$

$$(C_{33} - P)(C_{11} + C_{12}) - 2(C_{13} + P)^2 > 0 \quad (1)$$

$$(C_{44} - P)(C_{11} - C_{12} - 2P) - 2C_{14}^2 > 0$$

**Table 2.** Elastic constants $C_{ij}$ in GPa, and Cauchy pressures $P_C$ in GPa under several pressures of $EuCd_2As_2$.

| Pressure (GPa) | $C_{11} = C_{22}$ | $C_{12}$ | $C_{13} = C_{23}$ | $C_{33}$ | $C_{44} = C_{55}$ | $C_{66}$ | $(C_{13} - C_{44})$ | $(C_{12} - C_{66})$ |
|---|---|---|---|---|---|---|---|---|
| 0 | 96.64 | 30.12 | 28.21 | 75.63 | 28.27 | 33.26 | -0.06 | -3.14 |
| 2 | 109.65 | 38.38 | 35.78 | 85.64 | 31.06 | 35.64 | 4.72 | 2.74 |
| 4 | 121.50 | 45.42 | 43.33 | 94.96 | 33.27 | 38.04 | 10.06 | 7.38 |
| 6 | 131.09 | 54.53 | 50.95 | 104.07 | 34.30 | 38.28 | 16.65 | 16.25 |
| 8 | 140.65 | 61.56 | 58.07 | 112.23 | 34.97 | 39.55 | 23.10 | 22.01 |

As shown in **Table 2**, the calculated elastic constants satisfy the stability conditions for the selected pressures. The elastic constant $C_{11}$ characterizes a material's resistance to compression along the [100] direction, while $C_{33}$ describes resistance along the [001] direction. $EuCd_2As_2$ is less compressible along the [100] direction compared to the [001] direction, since $C_{11}$ is significantly larger than $C_{33}$ under all pressures. This indicates stronger atomic bonding along a direction. $C_{12}$, $C_{13}$, and $C_{23}$ are the elastic constants that describe how stress along one crystal axis produces strain along a different axis. Specifically, $C_{12}$ links stress in the a-direction to strain along the b-axis; $C_{13}$ links stress in the a-direction to strain along the c-axis; and $C_{23}$ links stress in the b-direction to strain along the c-axis. For $EuCd_2As_2$, $C_{12}$ is moderately greater than $C_{13}$ under all pressures. So, it is easy to produce deformation along a direction by applying strain along the b-direction compared to the c-direction. The elastic constant $C_{44}$ represents the material's ability to resist shear deformation when a tangential force is applied along the [010] direction on the (100) plane. For $EuCd_2As_2$, $C_{44}$ has a very low value compared to $C_{11}$ and $C_{33}$. So, it is easier to deform the material by applying a shearing force than a compressive force. The resistance of the (100) plane to shear along the [011] direction is characterized by the elastic constant $C_{55}$. From **Table 2,** we can see that there is a small difference between $C_{55}$ and $C_{66}$. With increasing pressure, both parameters increase slightly, but $C_{66}$ remains greater than $C_{55}$. From **Fig. 3 (b)** and **Table 2,** we can see that $C_{11}$, $C_{33}$, $C_{44}$, and $C_{66}$ increase with increasing pressure. $C_{12}$ and $C_{13}$ increase with increasing pressure. Therefore, we can conclude that pressure can be a useful tool to tune the elastic properties of $EuCd_2As_2$. Positive Cauchy pressure indicates ductility, while negative Cauchy pressure indicates a brittleness. Positive Cauchy pressure indicates predominance of ionic/metallic bonding where negative Cauchy pressure indicates predominance of covalent bonding. For trigonal $EuCd_2As_2$ the Cauchy pressures can be defined as ($C_{13} – C_{44}$) for (100) and (010) planes and ($C_{12} – C_{66}$) for the (001) plane [22]. From **Table 2,** we see that at zero pressure the Cauchy pressures are negative, which indicates a brittle nature. For higher pressures, the Cauchy pressures are positive, which indicates a ductile nature, and they increase, indicating that the ductility of the material can be adjusted by applied pressure.

### 3.2.2 Moduli of elasticity

To clearly understand the overall mechanical characteristics, the elastic moduli (Y, B, and G) are calculated. To calculate these elastic moduli, the Voigt-Reuss-Hill (VRH) method is used [23], [24], [25] and listed in **Table 3**. The Young's modulus for $EuCd_2As_2$ is found to be 74.13 GPa under zero pressure, indicating relatively low stiffness compared with many structural metals, such as iron. Young's modulus is a measure of resistance to longitudinal deformation. So this lower value of Young's modulus indicates the soft nature of this compound. The critical thermal shock coefficient is inversely proportional to Young's modulus [26]. So the compound in this study has a moderate critical thermal shock coefficient. From **Fig. 3 (c)** and **Table 3,** we can see that with increasing pressure from 0 to 2 GPa, the Young's modulus increases from 74.13 GPa to 81.06 GPa. Then, from 2 to 8 GPa pressure, it increases moderately up to 93.19 GPa. This increase indicates that $EuCd_2As_2$ becomes stiffer against elastic deformation as the applied pressure increases. The value of Young's modulus at 8 GPa of pressure is much lower than that of pure steel but is comparable to that of brass. The bulk modulus B describes the material's resistance to volume compression or fracture under applied pressure, whereas the shear modulus G characterizes its resistance to shape deformation or plastic distortion. The bulk modulus is closely related to the average binding energy of the atoms in crystals. For this compound, the bulk modulus is greater than the shear modulus at zero pressure, which indicates that the shear component should control the mechanical failure of $EuCd_2As_2$. From **Table 3** and **Fig. 3 (c),** we can see bulk modulus and shear modulus also increase with increasing pressure. **Table 3** further shows that the difference between the Voigt and Reuss values is very small. According to Hill, when these upper and lower bounds are close to each other, it means that the material has low elastic anisotropy [25].

**Table 3.** Calculated values of bulk modulus (B) in GPa, shear modulus (G) in GPa and Young's modulus (Y) in GPa, Pugh ratio (G/B), Poisson's ratio ($\sigma$) of $EuCd_2As_2$ calculated in this study.

| Pressure (GPa) | $B_R$ | $B_v$ | $B_H = B$ | $G_R$ | $G_v$ | $G_H = G$ | G/B | $Y_{Hill}$ | $\sigma$ |
|---|---|---|---|---|---|---|---|---|---|
| 0 | 48.40 | 49.11 | 48.76 | 29.35 | 30.12 | 29.73 | 0.61 | 74.13 | 0.25 |
| 2 | 57.42 | 58.31 | 57.87 | 31.44 | 32.55 | 31.99 | 0.55 | 81.06 | 0.27 |
| 4 | 65.91 | 66.90 | 66.41 | 33.02 | 34.64 | 33.83 | 0.51 | 86.76 | 0.28 |
| 6 | 74.36 | 75.46 | 74.91 | 34.07 | 35.37 | 34.72 | 0.46 | 90.22 | 0.30 |
| 8 | 82.05 | 83.21 | 82.63 | 34.74 | 36.29 | 35.51 | 0.43 | 93.19 | 0.31 |

### 3.2.3 Ductility and brittleness

The Pugh ratio is a parameter commonly used to assess the brittle or ductile nature of a material. It is defined as the ratio of the shear modulus to the bulk modulus, (G/B) [27]. A Pugh ratio below 0.57 generally indicates ductile behaviour, whereas a value above this threshold suggests brittle behaviour. As shown in **Table 3**, the Pugh ratio is below 0.57 at all investigated external pressures, indicating that $EuCd_2As_2$ exhibits ductile behavior under external pressure. Similarly, a Poisson's ratio greater than 0.26 is generally associated with ductility, whereas a lower value indicates a tendency towards brittle behaviour [28]. **Table 3** shows that the Poisson's ratio of $EuCd_2As_2$ is 0.25 at zero pressure, suggesting that the material is close to the conventional ductile-brittle threshold. Poisson's ratio can also provide insight into the bonding characteristics of a material. Values between 0.25 and 0.50 are commonly associated with the dominance of central forces in stabilizing the material [29], [30]. whereas values outside this range may suggest a greater contribution from non-central forces. Since the Poisson's ratio of $EuCd_2As_2$ falls within this range over the investigated pressure interval, the results suggest that central forces play an important role in its mechanical stability. As the pressure increases from 0 to 2 GPa, the Poisson's ratio rises from 0.25 to 0.27, while the Pugh ratio decreases from 0.61 to 0.55. From 2 to 8 GPa, the Pugh ratio continues to decrease, whereas the Poisson's ratio increases further, as shown in **Fig. 4**. These trends indicate that external pressure can be used to regulate the ductile-brittle response of $EuCd_2As_2$

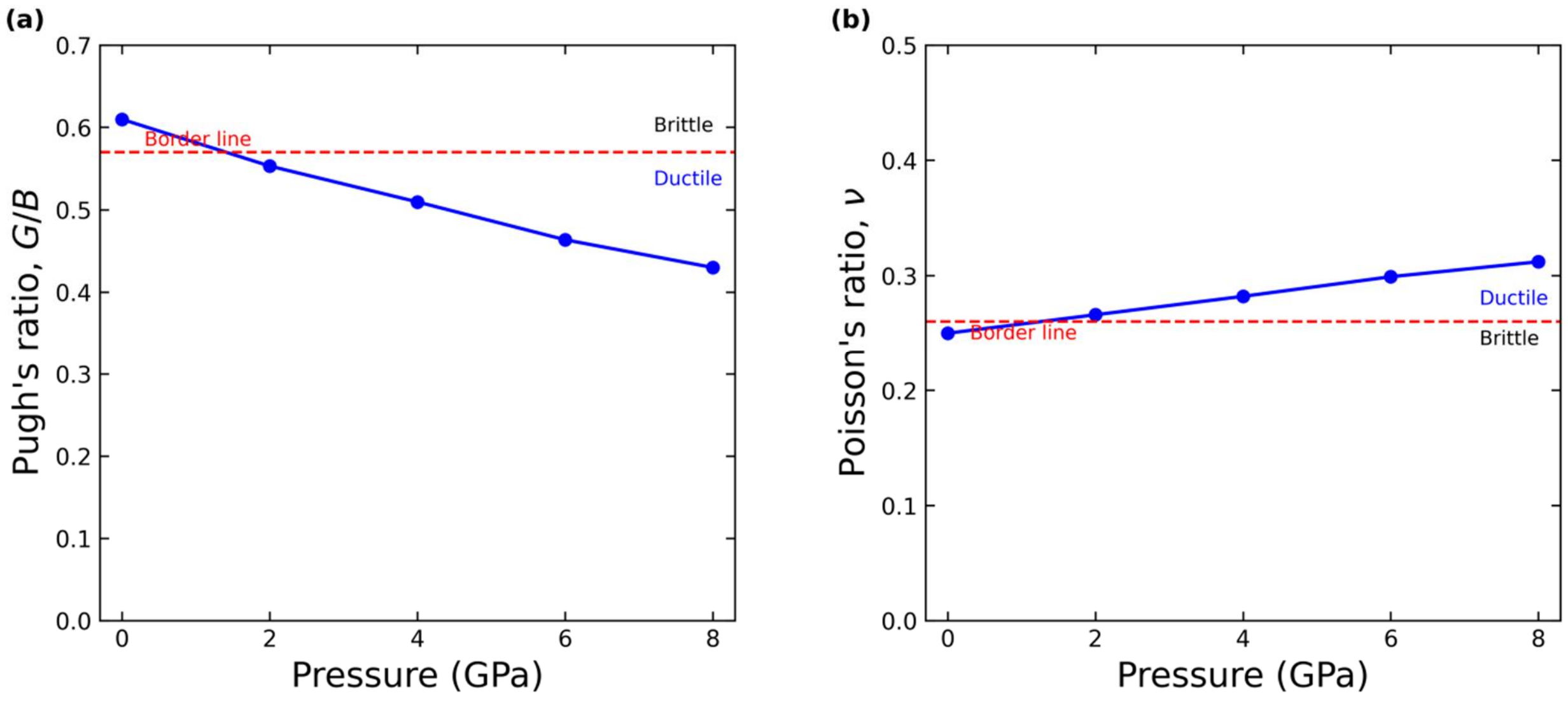

**Figure 4.** (a) Pugh's ratio vs applied pressure in GPa (b) Poisson's ratio $\nu$ vs applied pressure in GPa.

### 3.2.4 Machinability

The machinability index ($\mu_M$) is defined by the following equation [31].

$$\mu_M = \frac{B}{C_{44}} \quad (2)$$

This parameter indicates how easily a material can be cut, drilled, milled, or shaped using machine tools. In this study, the machinability index of $EuCd_2As_2$ is calculated to be 1.72 at zero pressure, suggesting moderate machinability. As shown in **Table 4**, the machinability index increases slightly to 1.86 at 2 GPa. It further increases to 2.00 at 4 GPa and reaches 2.36 at 8 GPa. These results indicate that the machinability of $EuCd_2As_2$ improves significantly with increasing pressure, suggesting that external pressure may provide a means of tuning its machinability.

**Table 4.** Machinability index at different pressures of $EuCd_2As_2$.

| Applied pressure in GPa | Machinability index ($\mu_M$) |
|---|---|
| 0 | 1.72 |
| 2 | 1.86 |
| 4 | 2.00 |
| 6 | 2.18 |
| 8 | 2.36 |

### 3.2.5 Elastic anisotropy

Elastic anisotropy is an important factor that affects how a material responds to different kinds of stress and strain. Anisotropy indices show how the material's mechanical properties vary across different directions. Elastic anisotropy influences many physical processes, such as how crystals deform, tiny cracks spread, quantum dots align, heat moves through the material, and overall defects dynamics. It also affects the material's toughness. In crystals, covalent bonds usually cause anisotropic (direction-dependent) behavior, while metallic bonds usually make the material behave isotropically (same in all directions) [20], [32]. The amount of anisotropy in atomic bonds on different crystal planes can be measured using shear anisotropy factors $A_1$, $A_2$, and $A_3$. $A_1$, $A_2$, and $A_3$ are the shear anisotropy factors for the (100), (010), and (001) planes, measured along the [011]–[010], [101]–[001], and [110]–[010] directions, respectively [33]. If a crystal is isotropic with respect to the different shear planes, $A_1 = A_2 = A_3 = 1$. From **Table 5,** it is seen that both $A_1$ and $A_2$ have a value of 0.976 at zero pressure, but $A_3$ has a value very equal to 1 at zero pressure. With increasing pressure, $A_1$ and $A_2$ are approximately equal to 1, and $A_3$ is 1. So under ambient and higher pressures, the crystal is isotropic with respect to the different shear planes. The other anisotropy parameters $A_G$, $A_B$, $A_L$, $A^U$, and $A_{eq}$ are calculated using the formulae stated in references [34], [35], [36], [37]. When $A_G = A_B = 0$, the material exhibits perfect elastic isotropy. On the other hand, the condition $A_G = A_B = 1$ corresponds to the highest possible elastic anisotropy. In the present study, both $A_G$ and $A_B$ were found to have values very close to zero under all pressures, which indicates excellent elastic isotropy in compressibility and in shear. The universal anisotropy index $A^U$ considers contributions from both the shear and bulk moduli. If $A^U$ deviates from zero, the crystal exhibits elastic anisotropy. In contrast, $A^U = 0$ corresponds to a completely isotropic crystal [38]. In this study, $A^U$ was found to be equal to 0.145, which is small and indicates low anisotropy. But at the higher pressures values of $A^U$ lie within the range from 0.192 to 0.261. This indicates a small increment of anisotropy. In locally isotropic crystals, $A_{eq}$ is equal to 1. But in this study, $A_{eq}$ was found to have a value of 1.413 at zero pressure, which indicates a small anisotropy. The parameter $A_L$ generally varies within the interval 0-10.26 for inorganic crystalline compounds [39]. $A_L = 0$ corresponds to complete elastic isotropy. The computed value of $A_L$ at zero pressure is 0.041, which is a further indication of very low anisotropy. At higher pressures, the value of $A_L$ is almost zero. From **Fig. 5 (a)**, it is seen that with an increment in pressure from 0 to 6 GPa, there is an increment in $A_1$ and in $A_2$, from 6 to 8 GPa $A_1$ and in $A_2$ decreases. Pressure has noticeable effect on $A_{eq}$, and $A^U$ as seen in **Fig. 5 (b)**.

In an isotropic solid, the three-dimensional representations of Young's modulus (Y), shear modulus (G), linear compressibility $\beta$, and Poisson's ratio ($\sigma$) are spherical, indicating that these properties are independent of direction. Any deviation from a spherical shape reflects elastic anisotropy. In this study, the directional dependence of these properties for $EuCd_2As_2$ was analyzed using the ELATE tool [40]. The elastic stiffness matrix was used as input to ELATE to generate the three-dimensional surfaces shown in **Figs. 6-9**, allowing the anisotropic behavior of Young's modulus, shear modulus, Poisson's ratio, and linear compressibility to be examined.

**Table 5.** Shear anisotropy factor ($A_1$, $A_2$ and $A_3$ ), the universal anisotropy index $A^U$ , equivalent Zener anisotropy measure $A_{eq}$ , anisotropy in shear $A_G$ , anisotropy in compressibility $A_B$ , universal log-Euclidean index $A_L$ , of $EuCd_2As_2$.

| Pressure (GPa) | $A_1$ | $A_2$ | $A_3$ | $A^U$ | $A_{eq}$ | $A_G$ | $A_B$ | $A_L$ |
|---|---|---|---|---|---|---|---|---|
| 0 | 0.976 | 0.976 | 1.000 | 0.145 | 1.413 | 0.013 | 0.007 | 0.041 |
| 2 | 1.004 | 1.004 | 1.000 | 0.192 | 1.488 | 0.017 | 0.008 | 0.034 |
| 4 | 1.025 | 1.025 | 1.000 | 0.261 | 1.588 | 0.024 | 0.007 | 0.032 |
| 6 | 1.030 | 1.030 | 1.000 | 0.205 | 1.507 | 0.019 | 0.007 | 0.024 |
| 8 | 1.023 | 1.023 | 1.000 | 0.237 | 1.554 | 0.022 | 0.007 | 0.028 |

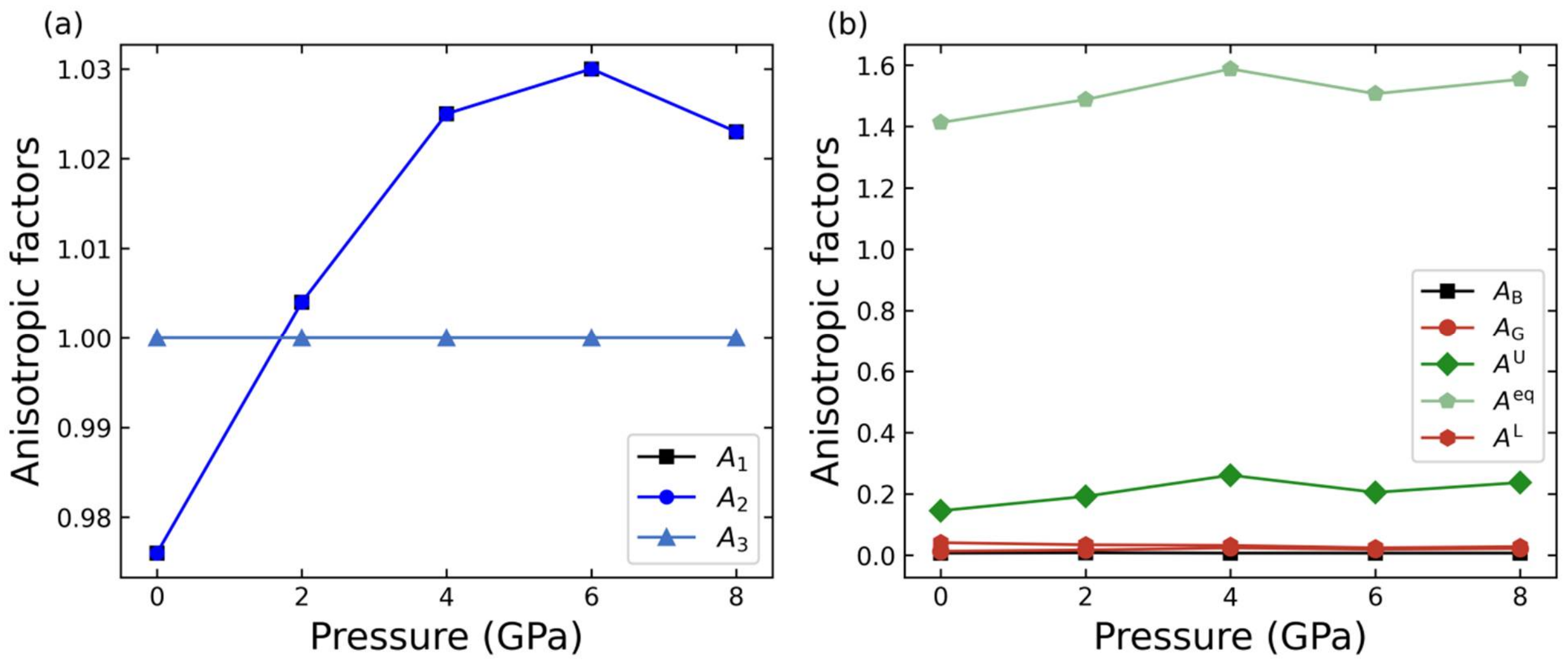

**Figure 5.** Elastic anisotropy indices and their pressure dependence for $EuCd_2As_2$.

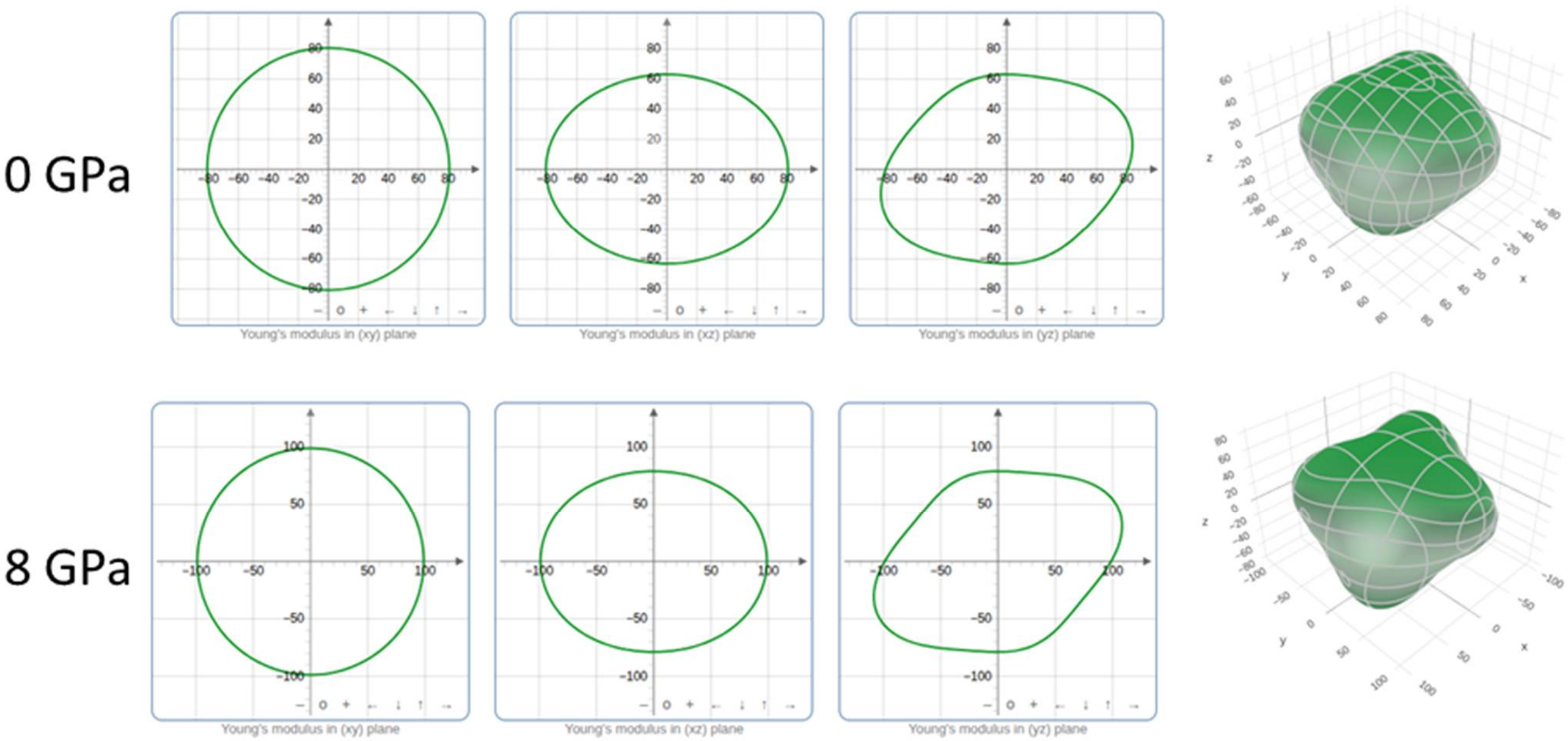

**Figure 6.** Directional dependence in Young's modulus (Y) of $EuCd_2As_2$ compound.

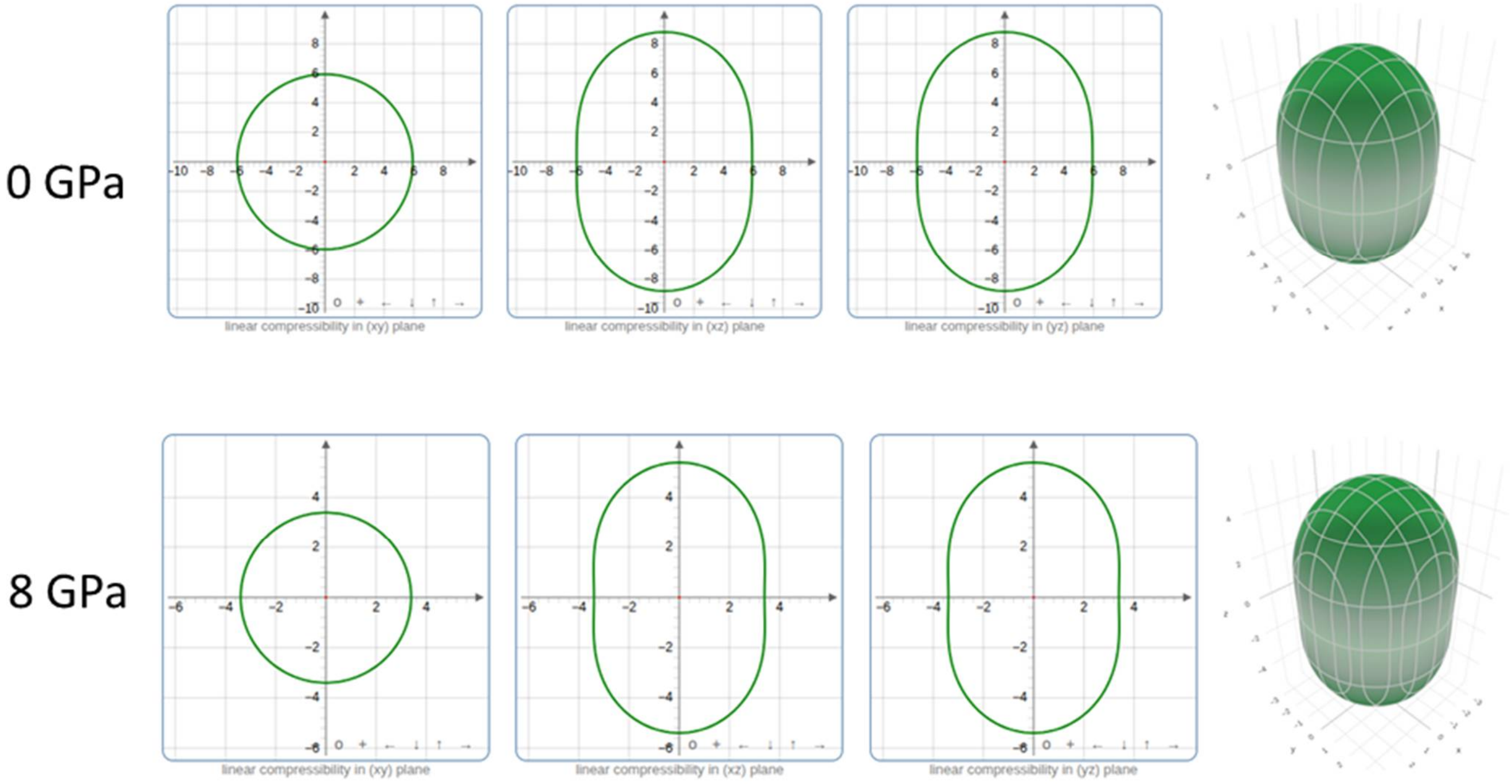

**Figure 7.** Directional dependence in linear compressibility $(\beta)$ of $EuCd_2As_2$ compound.

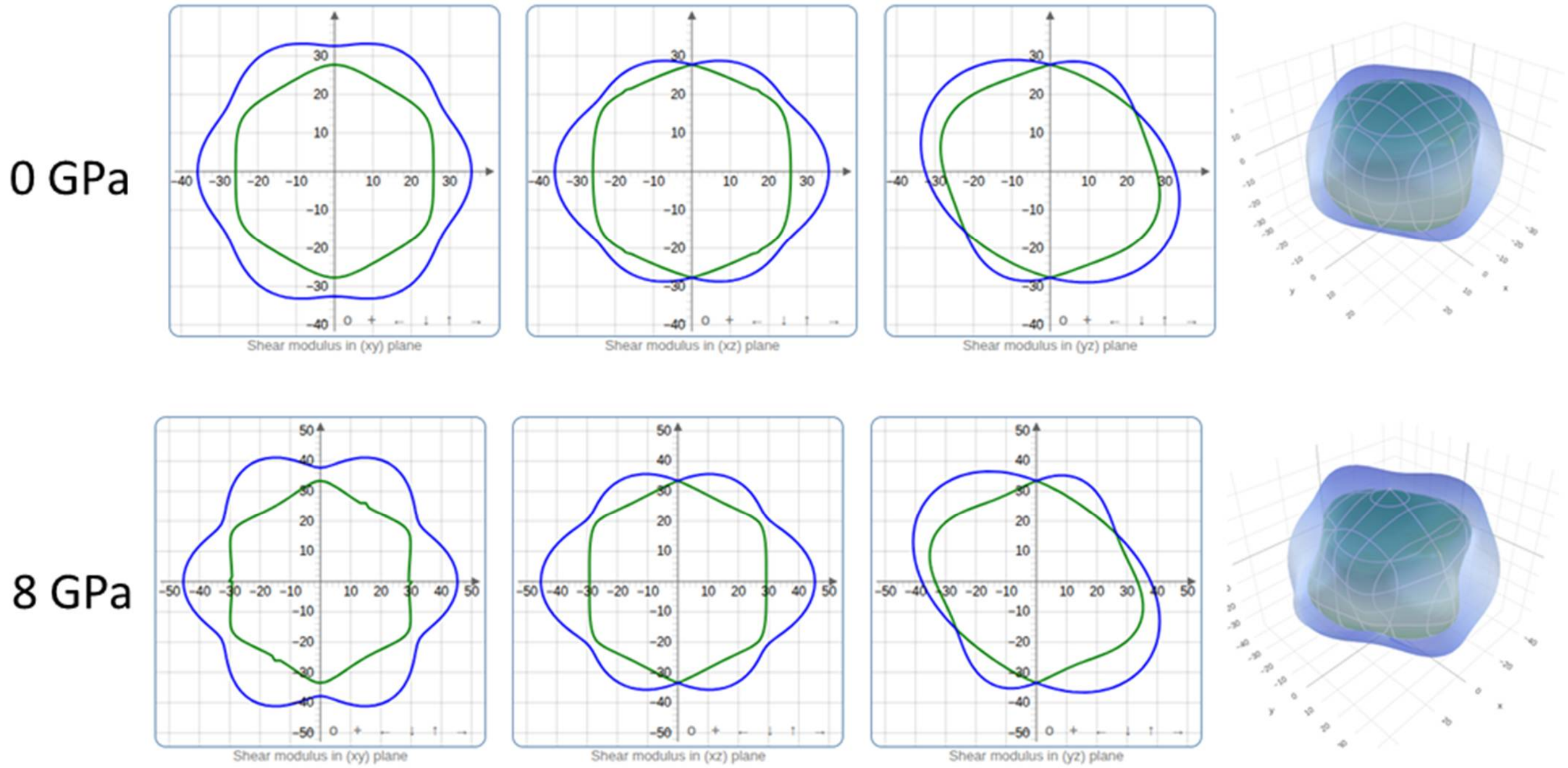

**Figure 8.** Directional dependence in shear modulus (G) of $EuCd_2As_2$ compound.

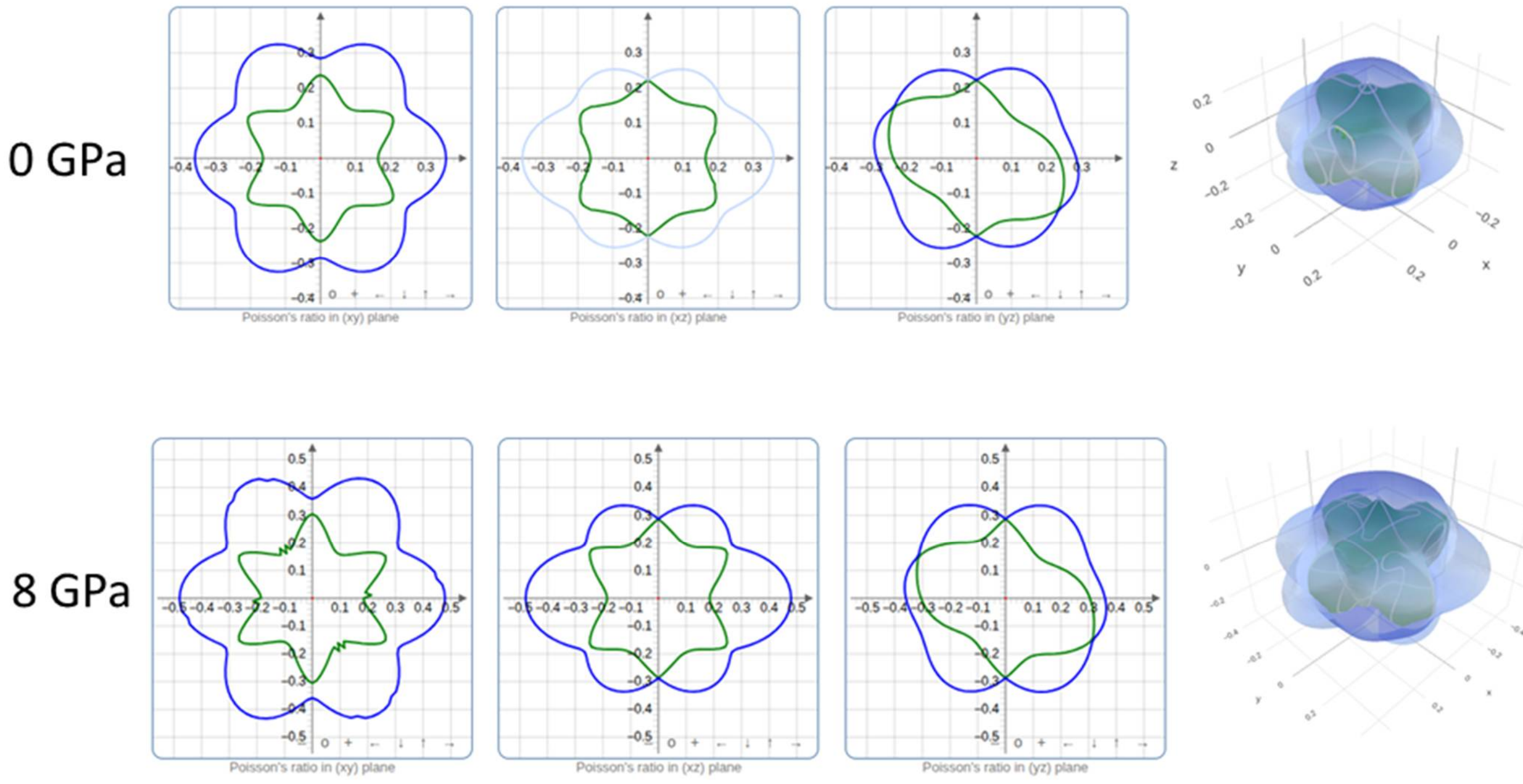

**Figure 9.** Directional dependence in Poisson's ratio ($\sigma$) of $EuCd_2As_2$ compound.

### 3.3 Thermophysical properties

### 3.3.1 Debye temperature

The Debye temperature $\Theta_D$ is an important parameter, defined as the temperature above which the phonon wavelength in a material is nearly equal to the interatomic separation. This value is a critical threshold for distinguishing between high and low temperature regimes of a solid. Understanding of Debye temperature ($\Theta_D$) is essential for characterizing various thermophysical properties, including melting temperatures, bonding forces, thermal conductivity, specific heat, and phonon dynamics. In the high temperature limit (T > $\Theta_D$), all vibrational modes have approximately equal energy ($k_B T$). At temperatures below $\Theta_D$, high-frequency modes are effectively "frozen," and only the acoustic modes are responsible for vibrational excitations. As a result, the Debye temperature calculated from elastic constants aligns closely with values derived from specific heat measurements in the low-temperature range. In this study, the Debye temperature of $EuCd_2As_2$ is calculated using the Anderson model. Equation for Debye temperature is given by [41].

$$\Theta_D = \frac{h}{k_B}\left[\frac{3n}{4\pi}\left(\frac{N_A\rho}{M}\right)\right]^{\frac{1}{3}} v_m \tag{3}$$

In this equation $k_B$ is the Boltzmann constant, $h$ is Planck's constant, and $v_m$ is the mean sound velocity. The Debye temperature $\Theta_D$ for the compound $EuCd_2As_2$ is found to be 233.29 K at zero pressure. This value is low, indicating that this compound is not stiff in nature, consistent with the elastic parameter analysis. But with increasing pressure, the Debye temperature rises to 240.88 K for 2 GPa, and it is 246.54 K for 4 GPa, and finally at 8 GPa it is 251.44 K , as shown in Table 6 and Fig. 10 (b). Thus, pressure is an effective parameter to control a large number of thermophysical properties of this compound.

**Table 6.** Pressure dependent Debye temperature of $EuCd_2As_2$.

| Applied pressure in GPa | Debye temperature in K |
|---|---|
| 0 | 233.29 |
| 2 | 240.88 |
| 4 | 246.54 |
| 6 | 249.39 |
| 8 | 251.44 |

### 3.3.2 Melting temperature

Another important thermophysical parameter of solids to understand stronger atomic interaction, bonding energy, cohesive energy, and coefficient of thermal expansion is the melting temperature, $T_m$. The following empirical relationship is used to calculate the melting temperature in terms of the elastic constants $C_{11}$ and $C_{33}$ [42]**:**

$$T_m = 354 + 1.5\,(2C_{11} + C_{33}) \quad (4)$$

The estimated melting temperature was found to be 757.36 K at zero pressure. This melting temperature is much lower in comparison to carbon steel [43]. So, the compound $EuCd_2As_2$ is suitable only for low temperature applications. From **Fig. 10 (c)**, it is evident that there is a notable effect of pressure on melting temperature. Therefore, by varying the pressure, atomic interaction, bonding energy, cohesive energy, and coefficient of thermal expansion can be tuned. Moreover, with increasing pressure, the thermal stability can be improved for $EuCd_2As_2$.

**Table 7.** Pressure dependent melting temperature for $EuCd_2As_2$.

| Pressure in GPa | Melting temperature in K |
|---|---|
| 0 | 757.36 |
| 2 | 811.44 |
| 4 | 860.93 |
| 6 | 903.38 |
| 8 | 944.30 |

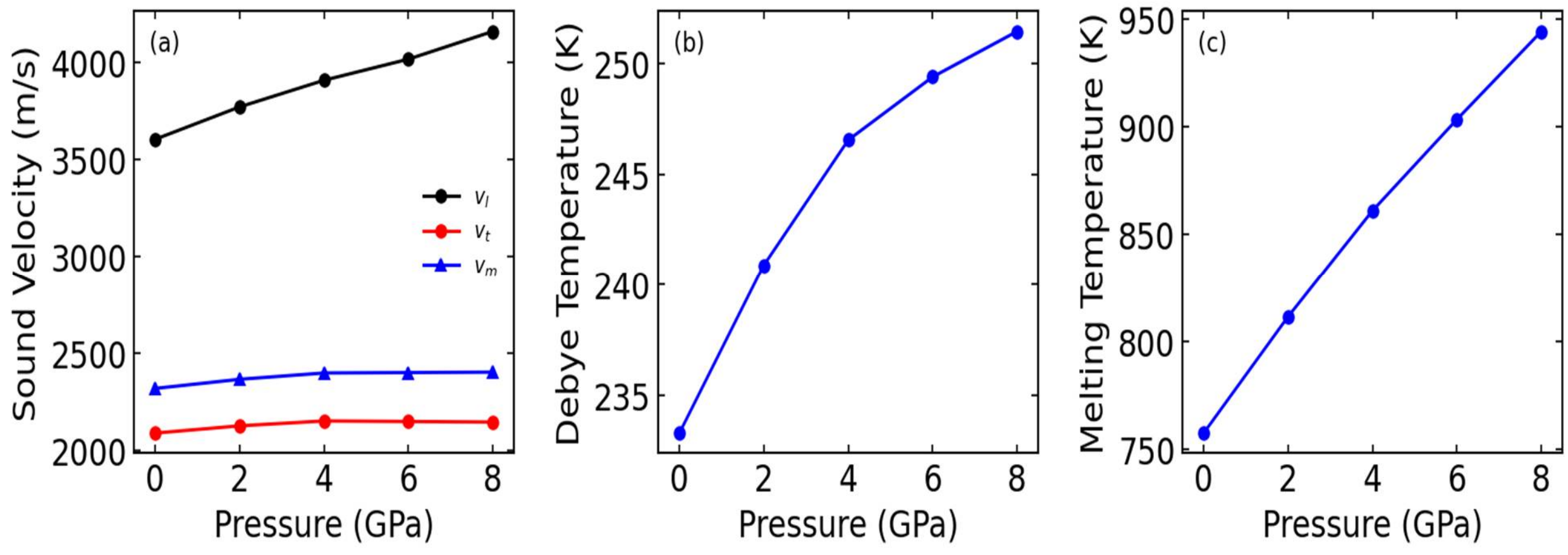

**Figure 10.** Pressure dependent (a) sound velocities, (b) Debye temperature, (c) melting temperature.

### 3.3.3 Grüneisen parameter and phonon thermal conductivity

The Grüneisen parameter, denoted by γ, is an important thermodynamic quantity that measures the degree of anharmonicity in a solid. It plays a significant role in several physical properties, such as thermal expansion, thermal conductivity, acoustic wave attenuation, and the temperature dependence of elastic constants. A larger value of γ indicates stronger anharmonic effects within the material. Using Poisson's ratio (σ), the Grüneisen parameter for $EuCd_2As_2$ can be estimated from the following relation [44]:

$$\gamma = \frac{3(1+\sigma)}{2(2-3\sigma)} \tag{5}$$

From **Table 8** and **Fig. 11 (a)**, it is seen that the Grüneisen parameter increases with pressure with applied pressure.

**Table 8.** Pressure-dependent Grüneisen parameter $\gamma$ and phonon thermal conductivity $k_{ph}$ in W/mK at 300 K.

| Applied pressure GPa | Grüneisen parameter $\gamma$ | Phonon thermal conductivity $k_{ph}$ |
|---|---|---|
| 0 | 1.50 | 8.702 |
| 2 | 1.60 | 7.825 |
| 4 | 1.67 | 7.560 |
| 6 | 1.77 | 6.422 |
| 8 | 1.84 | 5.948 |

Phonon thermal conductivity quantifies heat transfer through the vibrational motion of lattice ions in a solid [44], [45]. This parameter is vital for determining the energy conversion efficiency of thermoelectric materials. While both electrons and phonons act as heat carriers, electrons are the dominant transporters in metals at low temperatures. However, for materials used in high-temperature applications such as heat sinks, sensors, and thermal barrier coatings, study of lattice thermal conductivity is essential. Using the Slack model, the lattice thermal conductivity ($k_{ph}$) is estimated by the following expression [44]:

$$k_{ph} = A(\gamma)\frac{M_{av}\Theta_D^3\delta}{\gamma^2 n^{\frac{2}{3}} T} \tag{6}$$

Here, $M_{av}$ (in kg/mol) is the average atomic mass, δ is the cubic root of the average atomic volume in meter, and $\gamma$ is the Grüneisen constant, and the factor A ($\gamma$) can be calculated using the following equation [46]:

$$A(\gamma)\frac{4.85628 \times 10^{7}}{2(1-\frac{0.514}{\gamma}-\frac{0.228}{\gamma^{2}})} \tag{7}$$

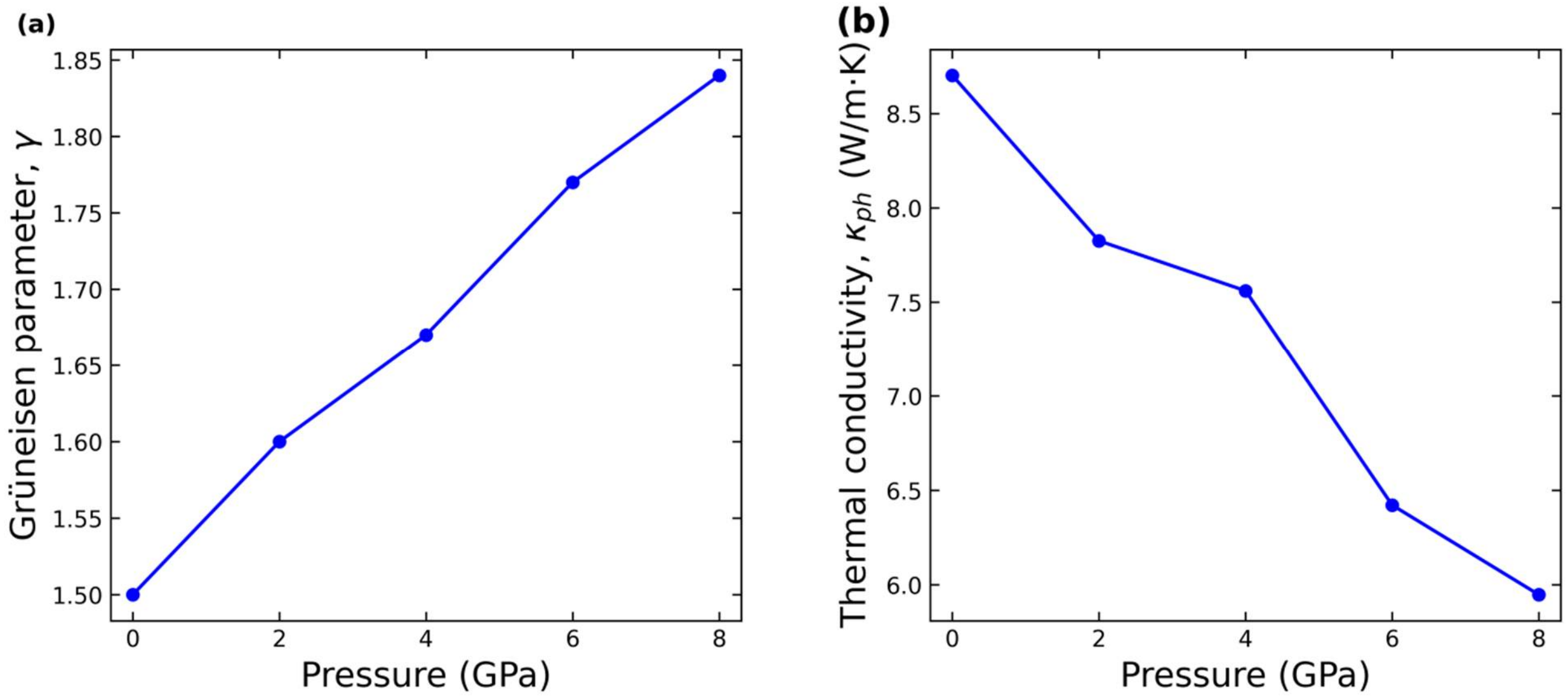

**Figure 11.** Pressure dependent (a) Grüneisen constant, and (b) phonon thermal conductivity at 300 K.

It is observed that $EuCd_2As_2$ possesses a phonon thermal conductivity 8.702 W/m.K at room temperature at zero pressure (**Table 8**). As we increase pressure, the phonon thermal conductivity decreases to 7.825 W/m.K at 2 GPa, and then it decreases and becomes 7.560 W/m.K at 4 GPa. It further decreases and finally becomes 5.948 W/m.K at 8 GPa. So, the phonon thermal conductivity can be adjusted over a wide range by changing pressure as seen from the **Fig. 11 (b)**.

### 3.4 Acoustic velocities

The speed of sound in a material is related to stiffness, crystal density, Debye temperature and thermophysical properties [47]. It also conducts electricity and heat. In recent years, many fields such as physics, materials science, musical instrument design, seismology, geology, and medical science have become very interested in studying the acoustic properties of materials and composites. The bulk and shear moduli are related to the speed of longitudinal and transverse waves of sound by the following equations [48]:

$$v_l = \sqrt{\frac{3B + 4G}{3\rho}} \tag{8}$$

$$v_t = \sqrt{\frac{G}{\rho}} \tag{9}$$

Using the following equation, the average sound velocity in a solid can be determined from transverse and longitudinal sound velocities [48]:

$$v_m = \left[\frac{1}{3}\left(\frac{2}{v_t^3} + \frac{1}{v_l^3}\right)\right]^{-1/3} \qquad (10)$$

In **Table 9**, the calculated acoustic velocities are shown. Every atom within a solid exhibits three modes of vibration: a single longitudinal mode and two transverse modes. In an anisotropic crystal, pure longitudinal and transverse modes manifest only along specific axes. From **Table 9,** we can see that the longitudinal velocity $v_l$ is greater than the transverse velocity $v_t$ under all pressures. With an increasing pressure, from 0 to 8 GPa there is a noticeable increase in both $v_l$ and $v_t$ [**Fig. 10 (a)**]. This increment in sound velocities indicates that with applied pressure the hardness of the compound increases, which is consistent with the analysis in the previous section. In $EuCd_2As_2$, transverse and longitudinal modes can only be present in the symmetry directions of [100] and [001] because of the trigonal symmetry of the material. The acoustic velocities of $EuCd_2As_2$ in these primary directions can be computed using the single crystal elastic constants [49].

[100] direction:

$$[100]_{vl} = \sqrt{\frac{C_{11} - C_{12}}{2\rho}}; [010]_{vt1} = \sqrt{\frac{C_{11}}{\rho}}; [001]_{vt2} = \sqrt{\frac{C_{44}}{\rho}} \qquad (11)$$

[001] direction:

$$[001]_{vl} = \sqrt{\frac{C_{33}}{\rho}}; [100]_{vt1} = [010]_{vt2} = \sqrt{\frac{C_{44}}{\rho}} \qquad (12)$$

From **Table 10**, it is noticeable that there is a significant anisotropy in the longitudinal velocity along [100] and [001] directions. Anisotropy in sound velocity is caused by anisotropy in elastic properties.

**Table 9.** Calculated density in gm/cm$^3$, longitudinal, transverse and average sound velocities ($v_l$, $v_t$ and $v_m$ in m/s) of $EuCd_2As_2$.

| Pressure in GPa | $\rho$ | $v_l$ | $v_t$ | $v_m$ |
|---|---|---|---|---|
| 0 | 6.81 | 3602.86 | 2089.45 | 2318.76 |
| 2 | 7.08 | 3769.55 | 2126.69 | 2365.67 |
| 4 | 7.31 | 3906.60 | 2151.73 | 2398.12 |
| 6 | 7.52 | 4015.25 | 2149.00 | 2400.17 |
| 8 | 7.71 | 4106.00 | 2145.87 | 2402.59 |

**Table 10.** Anisotropic sound velocities(in m/s) in $EuCd_2As_2$ along principal crystallographic direction under several pressures.

| Applied pressure in GPa | [100] direction | | | [001] direction | | |
|---|---|---|---|---|---|---|
| | $[010]_{vt1}$ | $[100]_{vl}$ | $[001]_{vt2}$ | $[001]_{vl}$ | $[100]_{vt1}$ | $[010]_{vt2}$ |
| 0 | 2209.82 | 3766.96 | 2037.40 | 3332.43 | 2037.40 | 2037.40 |
| 2 | 2243.58 | 3935.51 | 2094.52 | 3478.02 | 2094.52 | 2094.52 |
| 4 | 2282.60 | 4079.60 | 2134.87 | 3607.13 | 2134.87 | 2134.87 |
| 6 | 2256.61 | 4175.90 | 2136.14 | 3720.58 | 2136.14 | 2136.14 |
| 8 | 2264.45 | 4270.46 | 2129.49 | 3814.65 | 2129.49 | 2129.49 |

### 3.5 Electronic properties

#### 3.5.1 Band structure

To understand the electronic, optical, and magnetic properties of a material at a microscopic scale, we need to study its electronic band structure [38] [50] [51]. From the band structure, we can calculate the effective masses of charge carriers. The band structure also determines how charges move through the material and how atoms bond with each other [38]. In this study, the band structure of $EuCd_2As_2$ is calculated along high symmetry path Γ-A-H-K-Γ-M-L-H in **Fig. 12** for zero pressure and in **Fig. 13** for 8 GPa. The bands below the Fermi level are the valence band; on the other hand, the bands above the Fermi level are the conduction band. The horizontal line placed at zero energy indicates the Fermi level.

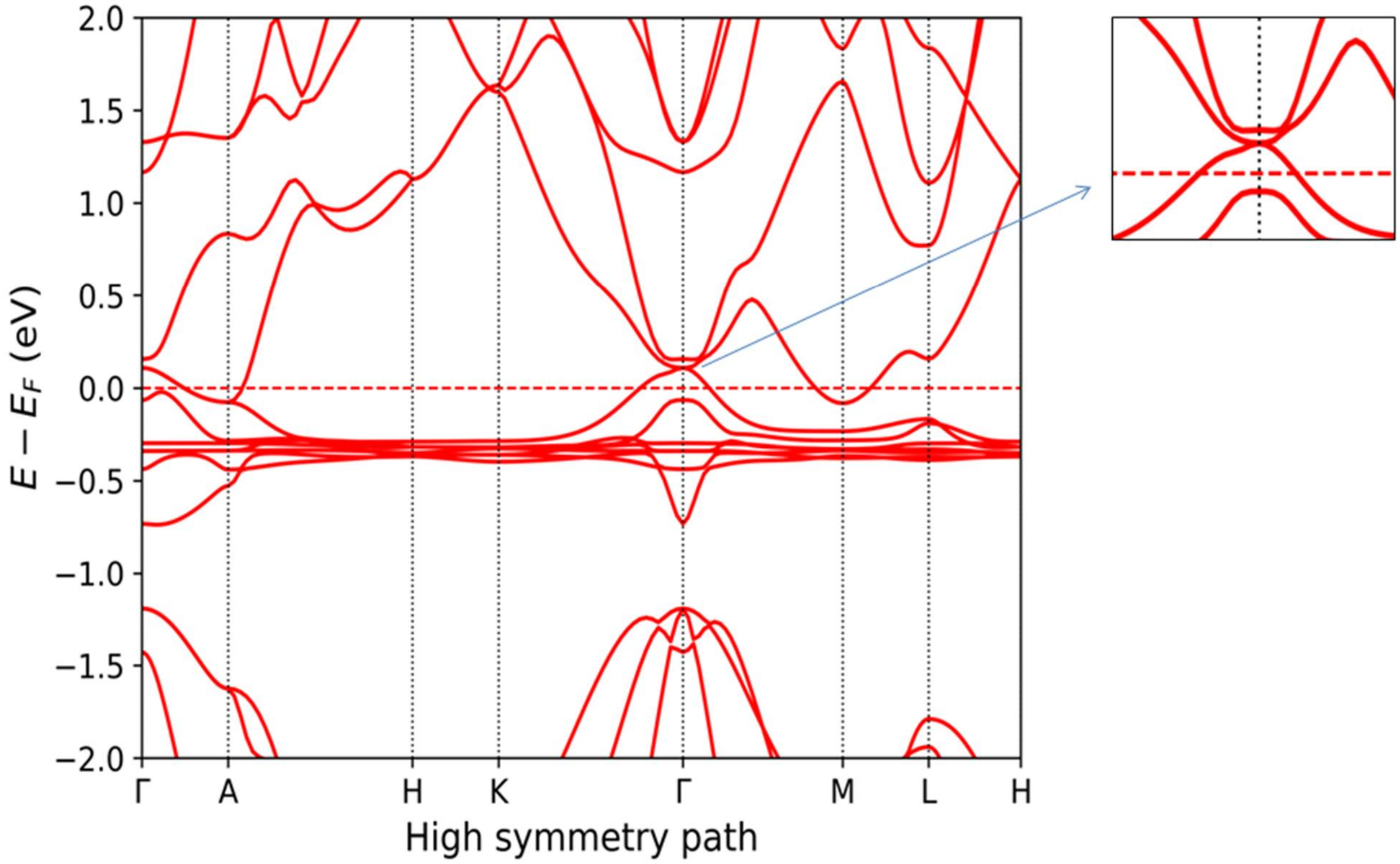

**Figure 12.** Band Structure obtained from Quantum Espresso in ferromagnetic phase at 0 GPa. A van Hove singularity close to the Fermi level in the conduction band is shown in the side panel.

Band structures of **Figs**. **12** & **13** are obtained from Quantum Espresso for the ferromagnetic phase. The effect of pressure is illustrated in **Fig. 4**. In **Fig. 15** for PDOS of Eu, we see that the PDOS for up spin and down spin are not symmetric so there is a nonzero net magnetic moment in $EuCd_2As_2$, which originates from f-electrons of Eu; this is consistent with the previous study [2]. Moreover, we found the magnetic moment to be 6.96 $\mu_B$ for 0 GPa. From **Fig. 12,** we can see that near the Fermi level, near Γ-point, there is no band gap, which indicates the semimetallic nature of the material in the ferromagnetic phase which is also consistent with the previous study [11]. At higher pressures, the semimetallic nature in the ferromagnetic phase (**Fig. 13**) is sustained. Interestingly, total number of bands crossing the Fermi level increases with increasing pressure. This is an indication of pressure induced Lifshitz transition in $EuCd_2As_2$. This should lead to pressure induced changes in the Fermi surface topology. Also, we find a van Hove singularity feature close to the Fermi level at the Γ-point. This feature disappears when pressure is raised (see **Fig. 13**). A major effect of pressure on the band structure is a consistent shift of dispersion branches to lower energies in the valence band and to higher energies in the conduction band. The band widths also increase with increasing pressure. In **Fig. 14,** we see the comparison in the band structure at 0 and 8 GPa. Overall, pressure affects the electronic band structure quite strongly. This suggests that pressure can be employed as an efficient tunning parameter to control the electronic properties of $EuCd_2As_2$.

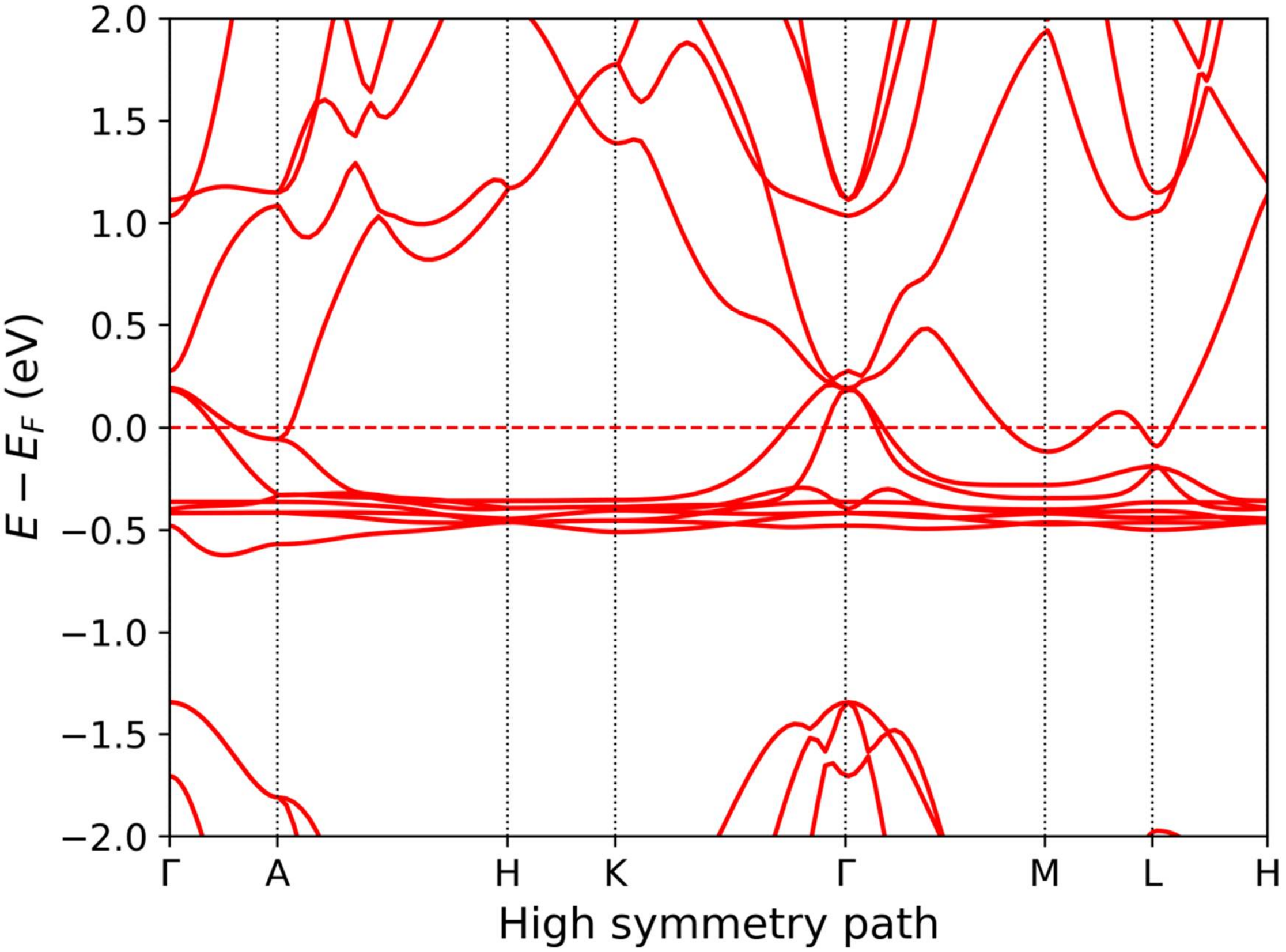

**Figure 13.** Band Structure obtained from Quantum Espresso in ferromagnetic phase at 8 GPa.

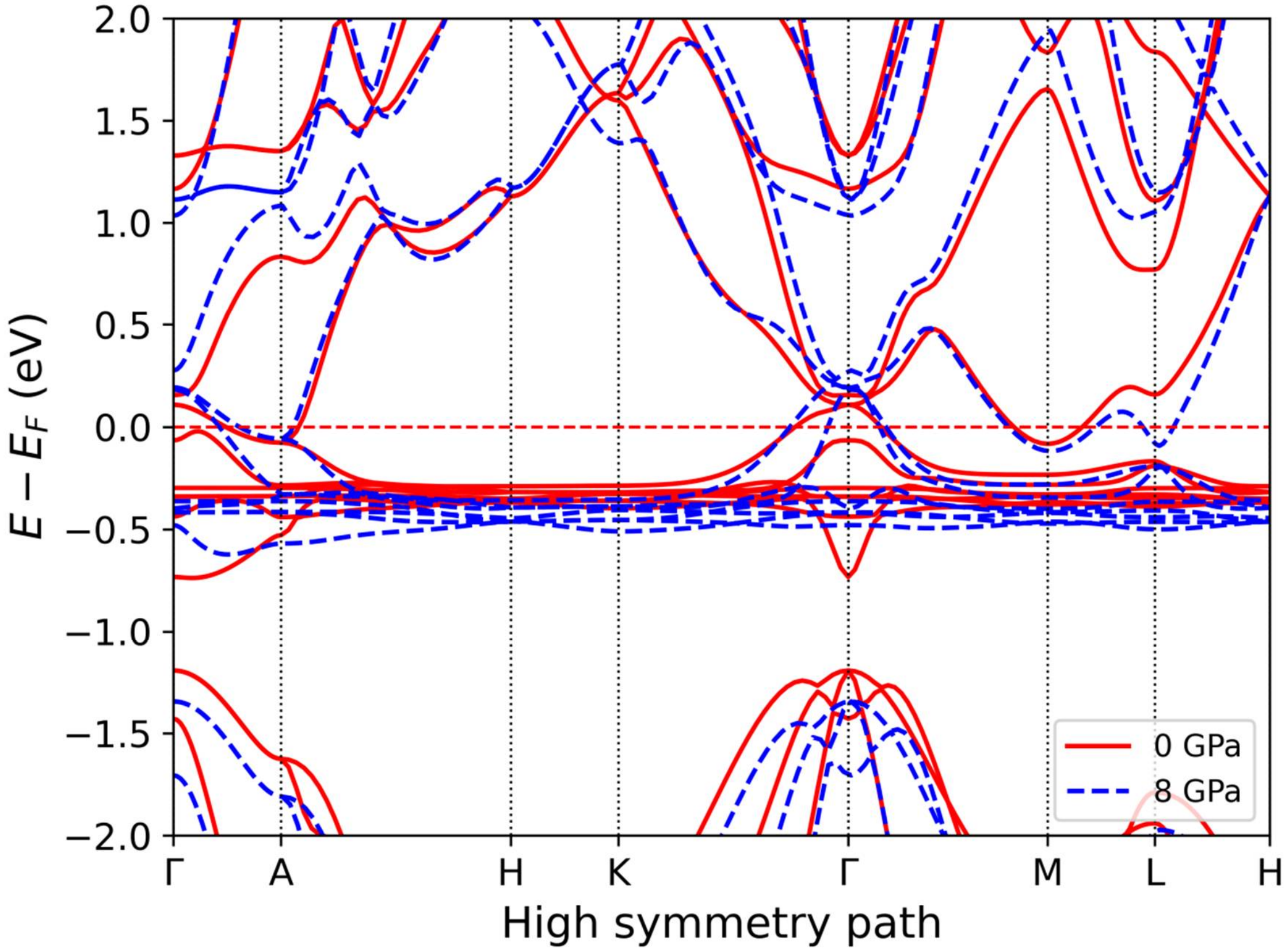

**Figure 14.** Band Structure obtained from Quantum Espresso in ferromagnetic phase at 0 and 8 GPa.

### 3.5.2 Electronic density of states

The density of states (DOS) describes the number of electronic states available per unit energy interval. The distribution of states in the valence and conduction bands provides useful insight into the electronic structure and can help explain the electrical, magnetic and optical properties of crystalline materials. The contributions of individual atoms and their orbitals can be investigated by analyzing the total density of states (TDOS) and partial density of states (PDOS). In this study, the TDOS and PDOS of $EuCd_2As_2$ at zero pressure are presented in **Fig. 15**. The relative positions and hybridization of atomic-orbital states can also provide insight into bonding and anti-bonding interactions. The vertical line in **Fig. 15** represents the Fermi level. The DOS exhibits prominent features at approximately −0.33 eV and 3.86 eV, which are attributed to bonding and anti-bonding-related states, respectively, based on the calculated electronic structure. The TDOS at the Fermi level is very small, suggesting a low density of available electronic states near the Fermi energy, consistent with semimetallic behavior supported by the band structure. A minimum in the TDOS between the prominent peaks, referred to as a pseudogap, lies close to the Fermi level. The presence of a pseudogap may indicate an energetically favorable distribution of electronic states; however, it does not, by itself, establish the thermodynamic stability of the compound.

In the valence-band region, As-p states make a major contribution to the PDOS of As, while Eu-f states dominate the Eu contribution. Cd-d states contribute little to the Cd PDOS near Fermi level. In the conduction-band region, the contribution from As-p states is comparatively small, whereas Eu-f and Eu-d states make noticeable contributions. A small contribution from Cd-p states is also observed. Near the Fermi level, the electronic states are contributed mainly by Eu-f, Eu-d, and As-p orbitals, according to the calculated PDOS. The peak in TDOS at about -10 eV comes from the contribution of Cd-d orbitals. The spin asymmetry in the PDOS arises solely from the Eu-f electrons and contributes to the ferromagnetic ground state. For the electronic orbitals involving As and Cd, the PDOS features are spin symmetric.

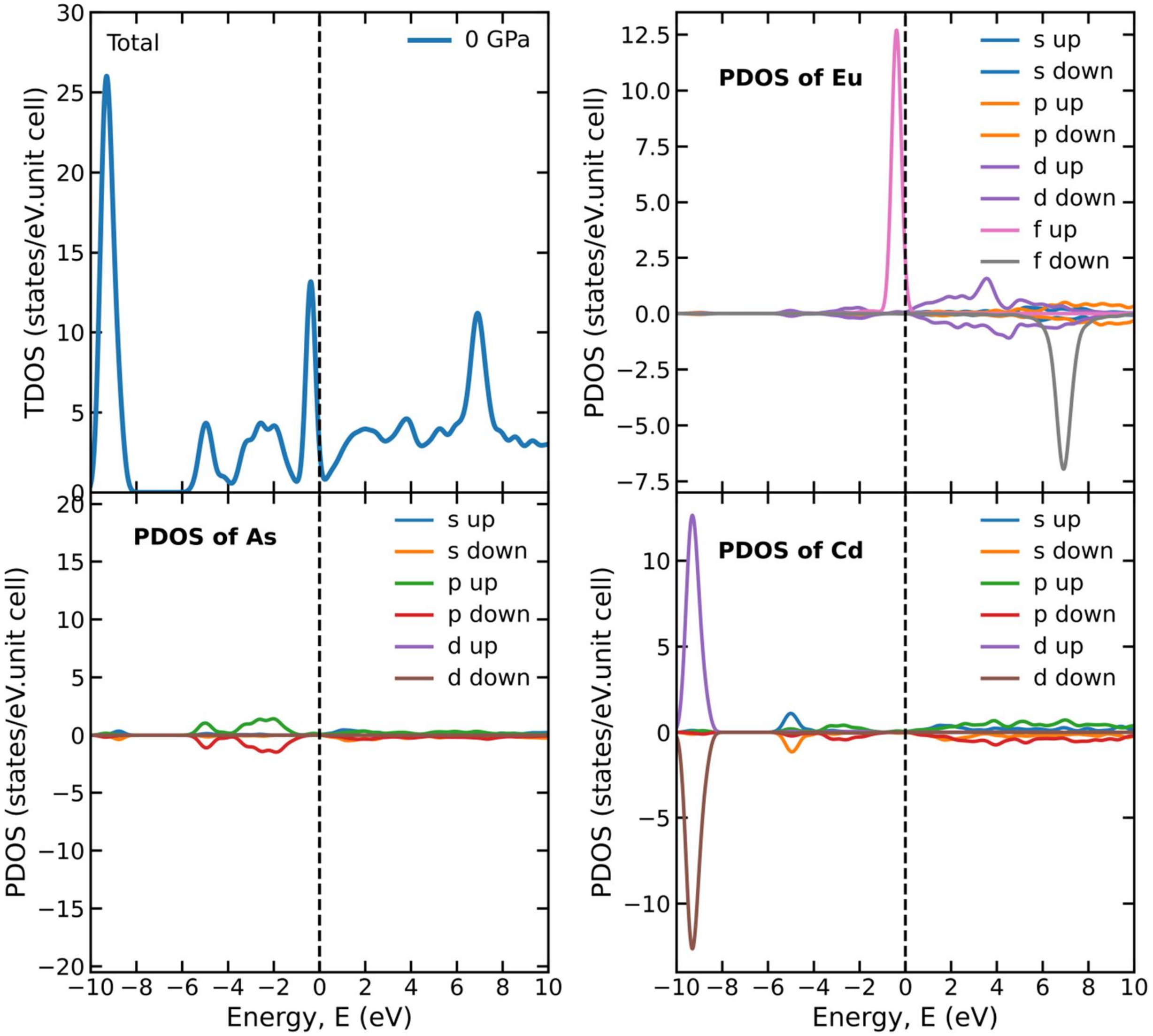

**Figure 15.** Total and partial density of states of $EuCd_2As_2$ at zero pressure. The vertical lines indicate the Fermi level.

### 3.6 Optical properties

Optical properties give information about how the material behaves upon incidence of an electromagnetic wave. Understanding the optical response in the visible-light region is particularly important for optoelectronic applications. The study of optical properties of solids has gained considerable attention because of their wide use in modern technologies such as display devices, sensors, lasers, photo detectors, photonics, solar cells, and photo-electrodes. Moreover, optical anisotropy must be considered, as many optical technologies such as displays, polarizers, and wave plates are based on this property. Different energy- or frequency-dependent optical properties, such as the dielectric function, loss function, refractive index, optical conductivity, absorption coefficient, and reflectivity, fully describe how a material responds to incident light [38], [53]. In this study, optical properties for $EuCd_2As_2$ are calculated for photon energies in the range from 0 to 30 eV, and the polarization for the electric field were taken to be in [100] and [001] directions. The computed optical parameters include the dielectric function ε(ω), the refractive index n(ω), the optical conductivity κ(ω), the reflectivity R(ω), the absorption coefficient α(ω), and the energy loss function L(ω). Here, $\omega = 2\pi f$ represents the angular frequency. In these calculations, a Drude damping (relaxation energy) value of 0.05 eV and a Gaussian smearing width of 0.5 eV were applied.

**Dielectric function:** In **Fig. 16 (a)**, both real parts and imaginary parts of the dielectric constant under 0, 4, and 8 GPa pressures are plotted. The real part explains the electrical polarization of materials upon incident radiation; on the other hand, the imaginary part shows how dielectric loss varies with the energy of incident radiation. The dielectric function approaches a very low value at about 9.80 eV of incident energy. Beyond about 3.55 eV, the dielectric constant goes to a negative value. So, there is a metallic signature as the dielectric function goes to zero from a negative value. Beyond about 14 eV, the imaginary part of the dielectric constant goes to zero, and the first peak in the loss function is found to be about 14.3 eV. Another noticeable feature is that for very low energy of incident wave, the real part of dielectric function for both polarizations have high value, and as the energy increases, the values of the real part of dielectric functions fall and approach zero beyond about 9.80 eV. The optical anisotropy in the dielectric spectra remains low at all pressures.

**Refractive Index:** The phase velocity of an electromagnetic wave inside a material can be determined by the real part of the refractive index, and how the electromagnetic wave (EMW) attenuates inside the material can be understood from the imaginary part of the refractive index (also known as excitation coefficient). At very low energy, the real part of refractive index has a high value of about 5.62 for both polarizations at every pressure, as shown in **Fig. 16 (b)**. This implies no significant optical anisotropy at low energies. There is no significant effect of pressure in this region. With increasing energy, values of the real part decrease and become very low beyond the energy of about 8.82 eV for all pressures. After 15 eV, the real part of the refractive index for both directions starts to increase slightly and reaches a maximum value at about 20.42 eV.

**Absorption coefficient:** In **Fig. 16 (c)**, the absorption coefficient is plotted for both [100] and [001] polarization directions. For both polarization directions, the absorption coefficient starts to increase from the 0 eV which indicates the semimetallic nature of the material. Two peaks are seen in **Fig. 16 (c)**; the first peak is observed at the energy of about 6.35eV. From 3.00 eV to 5.71 eV absorption coefficient is large. Its value in visible range is also significant. Thus, this compound is good at absorbing visible light as well as good at absorbing UV light. Beyond this energy, before the next peak, the absorption coefficient has a minimum value at an energy of about 18 eV (for both polarizations). The material is also a good absorber of high energy UV radiation. The absorption coefficient gets to a peak at an energy of about 21.80 eV. In this result, no significant optical anisotropy is observed. The high absorption coefficient in both visible and UV range indicates the potential of the material for optical application. There is significant effect of pressure on the absorption spectra. The compound becomes more absorbing as pressure increases and the peaks show blue shits with rising pressure.

**Optical conductivity:** The conduction of free charge carriers as a function of the photon energy can be learned from the optical conductivity of a material. This is the dynamic response of free charge carriers in materials, such as an electron or electron-hole pair (for materials with band gap). In **Fig. 16 (d)**, it is seen that for [100] polarization the conductivity starts from almost 0 eV for all pressures, indicating no band gap. Initially, optical conductivity increases and get the first peak within an energy range of 2.93 to 4.50 eV for different pressures. Beyond this range, optical conductivity decreases to a low value for both polarizations. Within the range from 4.29 eV to 9.0 eV, optical anisotropy is observed in optical conductivity.

**Reflectivity**: In **Fig. 16 (e)**, the reflectivity as a function of EMW energy is plotted for 0, 4, and 8 GPa. At zero energy and pressure, the reflectivity has the value of about 0.5. At this energy no significant anisotropy in the reflectivity spectra is observed. A peak in reflectivity is observed at about 7.41 eV. This peak is highly sensitive to pressure and grows in magnitude as pressure rises. In the visible and UV region, the compound is found to be a good reflector. At 15 eV, the reflectivity has a low value for both directions of polarization; the first peak in the loss function is also observed near this energy. Beyond about 18.50 eV, the reflectivity starts increasing and then decreases to a very small value at about 30 eV; the second peak in the loss function is observed at about 27.20 eV.

**Loss function**: The loss function shown in **Fig. 16 (f)** indicating how fast electron will lose energy moving through the material. A peak in the loss function is observed at an energy equivalent to the plasma frequency. In the loss function for zero pressure, we can find the first peak at about 14.30 eV and 13.78 eV for [001] and [100] polarizations, respectively. The second peak is observed at an energy of about 27.20 eV. These energies are equivalent to plasmon energies. The first peak in the loss function for [100] direction shifts slightly to the right with increasing pressure from 0 to 8 GPa, which is an indication of an increment of plasma frequency. Both pressure and electric field polarization have significant effect of the loss function profile.

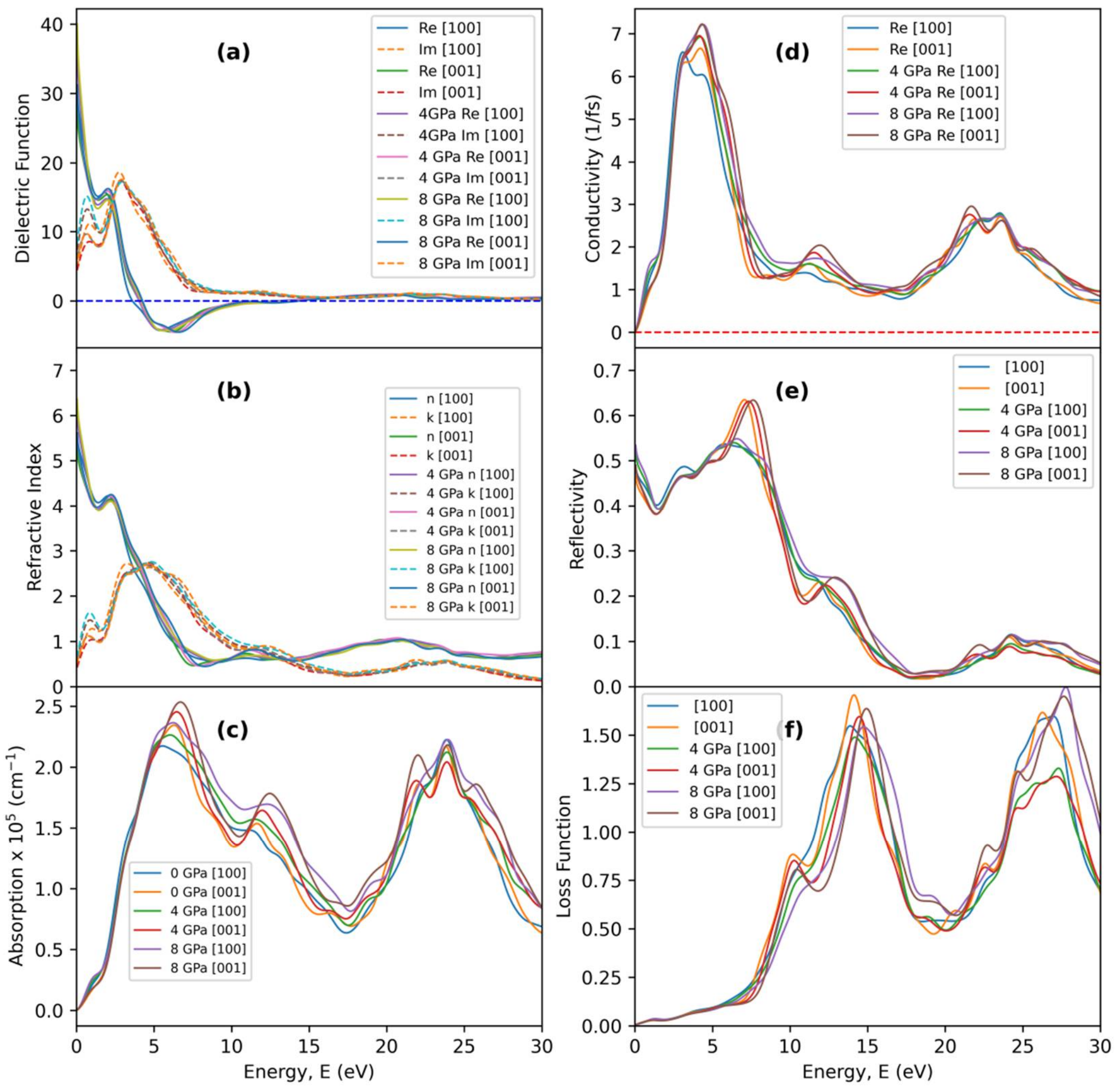

**Figure 16.** (a) Solid lines are for the real part of the dielectric function and dashed lines are the imaginary part of the dielectric function, (b) Solid lines are for the real part of the refractive index, and dashed lines are the imaginary part of the refractive index, (c) Absorption coefficient, (d) Optical conductivity, (e) Reflectivity, and (f) Loss function plot for 0, 4, and 8 GPa.

## 4. Conclusions

In this work, we performed a systematic pressure-dependent first-principles DFT study of the structural, elastic, thermophysical, acoustic, electronic and optical properties of $EuCd_2As_2$ using the GGA-PBE method. The calculated lattice parameters at ambient pressure are in good agreement with the experimentally obtained lattice parameters, confirming the reliability of our computational work. The elastic constants and the calculated mechanical properties reveal that $EuCd_2As_2$ meets the Born–Huang stability criterion at all pressures ranging from 0 to 8 GPa, thereby confirming its mechanical stability within this pressure range. The bulk modulus, shear modulus, and Young's modulus all increased with increasing pressure, indicating that the material progressively stiffened under pressure. At ambient pressure, the relatively low Young's modulus (74.13 GPa) reflects the soft, layered nature of the crystal structure. Both the Pugh and Poisson's ratios indicate that $EuCd_2As_2$ is brittle under zero pressure becomes ductile under pressures, and that the central interatomic forces govern its structural stability. The machinability index increases with applied pressure. This is a consequence of increased metallicity of the material under pressure. $EuCd_2As_2$ possesses small elastic and optical anisotropy. The Debye temperature increases from 233.29 K at 0 GPa to 251.44 K at 8 GPa, which is consistent with pressure-induced hardening. The melting temperature also increases with the applied pressure, from 757.36 K to 944.30 K, this indicates improved thermal stability under compression. The Grüneisen parameter increases with applied pressure. This indicates increment in the lattice anharmonicity. The phonon thermal conductivity decreases with applied pressure. All these demonstrate that pressure is an effective tool for tuning the thermophysical properties of this compound. Acoustic velocity calculations revealed significant anisotropy between the [100] and [001] crystallographic directions, both the longitudinal and transverse velocities increase with applied pressure. In the FM phase, the material remains semimetallic at all studied pressures from 0 to 8 GPa. The density of states analysis indicates that Eu-f, Eu-d, and As-p orbitals contribute strongly to the electronic states near the Fermi level. In the FM phase, the partial density of states of Eu shows an asymmetric distribution between spin-up and spin-down states. As a result, a magnetic moment of 6.96 $\mu_B$ per unit cell is obtained. A pseudogap is observed just above the Fermi level in the total density of states. Spin asymmetric density of states around the Fermi level suggests that $EuCd_2As_2$ has potential to be used in spintronics. Pressure has notable effect on the band structure and can be used to tune the electronic properties. For example, pressure changes the Fermi surface topology and removes the van Hove singularity feature as shown in **Fig. 14**. The optical properties of $EuCd_2As_2$ showed strong absorption within the visible and ultraviolet energy ranges. The high refractive index at low photon energies and broad optical conductivity further support the optoelectronic versatility of the material. The Reflectivity was good in the visible range, whereas from loss function plot plasmon resonances were found at approximately 13–15 eV and 27 eV.

All the physical properties studied are affected by pressure to varying degrees, with variations highly enhanced in the low pressure regime. Overall, this study indicates unambiguously that pressure is a powerful tool for controlling the mechanical, thermal, acoustic, and electronic properties of $EuCd_2As_2$. The pressure induced ductile nature, pressure tunable thermal conductivity, electronic band structure, and strong optical absorption and reflectivity of this compound across a wide energy range indicates that $EuCd_2As_2$ is a promising candidate for diverse technological applications.

## Data availability

The data sets generated and/or analyzed in this study are available from the corresponding author on reasonable request.

## Declaration of interest

The authors declare that they have no known competing financial interests or personal relationships that could have appeared to influence the work reported in this paper.

## Declaration of AI usage

The authors used Grammarly, ChatGPT (OpenAI) and Claude AI to improve the language, grammar, sentence structure, and readability of the manuscript. These tools were used solely for language polishing, and the authors take full responsibility for the content, interpretation, and conclusions presented in this paper

## Author Contributions

**Salauddin Muhammad Anis**: Methodology, Software, Formal Analysis, Writing- Original draft. **M. A. Afzal**: Methodology, Validation, Supervision, Writing-Reviewing and Editing. **R. S. Islam**: Validation, Supervision, Writing-Reviewing and Editing. **S. H. Naqib**: Conceptualization, Validation, Supervision, Writing- Reviewing and Editing.